\documentclass{aa}
\usepackage{graphicx}
\usepackage{hyperref}

\usepackage{xcolor}
\usepackage{txfonts}
\begin{document}

   \title{Three is the magic number - distance measurement of NGC 3147 using SN 2021hpr and its siblings}


   \author{B. Barna \inst{1} \and
          A. P. Nagy\inst{1} \and 
          Zs. Bora\inst{2,3} \and 
          D. R. Czavalinga\inst{4,5} \and 
          R. Könyves-Tóth\inst{1,6,7} \and
          T. Szalai\inst{1,4} \and 
          P. Székely\inst{1} \and 
          Sz. Zs{\'i}ros\inst{1} \and 
          D. B{\'a}nhidi\inst{5} \and 
          I. B. B{\'i}r{\'o} \inst{4, 5} \and
          I. Cs{\'a}nyi\inst{5} \and 
          L. Kriskovics\inst{2, 6} \and 
          A. P{\'a}l\inst{2, 6} \and 
          Zs. M. Szab{\'o}\inst{3, 6, 9, 10} \and 
          R. Szak{\'a}ts\inst{2, 6} \and 
          K. Vida\inst{2, 6} \and 
          Zs. Bodola\inst{1} \and 
          J. Vink{\'o}\inst{2, 6, 1, 8}
          }
          
   \institute{Department of Experimental Physics, Institute of Physics, University of Szeged, D{\'o}m t{\'e}r 9, 6720 Szeged, Hungary \\ 
              \email{bbarna@titan.physx.u-szeged.hu}
        \and 
        E{\"o}tv{\"o}s Lor{\'a}nd University, Department of Astronomy, P{\'a}zm{\'a}ny P{\'e}ter s{\'e}t{\'a}ny 1/A, 1117 Budapest, Hungary 
        \and
        Konkoly Observatory, Research Centre for Astronomy and Earth Sciences, Konkoly Th. M. {\'u}t 15-17., 1121 Budapest, Hungary 
         \and
         ELKH-SZTE Stellar Astrophysics Research Group, Szegedi {\'u}t, Kt. 766, 6500 Baja, Hungary 
         \and
         Baja Astronomical Observatory of University of Szeged, Szegedi {\'u}t, Kt. 766, 6500 Baja, Hungary 
        \and 
        CSFK, MTA Centre of Excellence, Konkoly Thege Mikl{\'o}s {\'u}t 15-17, 1121 Budapest, Hungary 
        \and 
        ELTE E{\"o}tv{\"o}s Lor{\'a}nd University, Gothard Astrophysical Observatory, 9400 Szombathely, Hungary 
        \and 
        E{\"o}tv{\"o}s Lor{\'a}nd University, Institute of Physics, P{\'a}zm{\'a}ny P{\'e}ter s{\'e}t{\'a}ny 1/A, 1117 Budapest, Hungary 
        \and 
        Max-Planck-Institut f{\"u}r Radioastronomie, Auf dem Hügel 69, 53121 Bonn, Germany 
        \and 
        Scottish Universities Physics Alliance (SUPA), School of Physics and Astronomy, University of St Andrews, North Haugh, St Andrews, KY16 9SS, UK 
             }

   \date{Received December 12 2022}

 
  \abstract
   {The nearby spiral galaxy NGC 3147 hosted three Type Ia supernovae (SNe Ia) in the past decades, which have been subjects of intense follow-up observations. Simultaneous analysis of their data provides a unique opportunity for testing the different light curve fitting methods and distance estimations.}
   {The detailed optical follow-up of SN 2021hpr allows us to revise the previous distance estimations to NGC 3147, and compare the widely used light curve fitting algorithms to each other. After the combination of the available and newly published data of SN 2021hpr, its physical properties can be also estimated with higher accuracy.}
   {We present and analyse new $BVgriz$ and Swift photometry of SN 2021hpr to constrain its general physical properties. Together with its siblings, SNe 1997bq and 2008fv, we cross-compare the individual distance estimates of these three SNe given by the SALT code, and also check their consistency with the results from the MLCS2k2 method. The early spectral series of SN 2021hpr are also fit with the radiative spectral code {\tt TARDIS} in order to verify the explosion properties and constrain the chemical distribution of the outer ejecta.}
   {After combining the distance estimates for the three SNe, the mean distance to their host galaxy, NGC 3127, is  $42.5 \pm 1.0$ Mpc, which matches with the distance inferred by the most up-to-date LC fitters, SALT3 and BayeSN. We confirm that SN~2021hpr is a Branch-normal Type Ia SN that ejected $\sim 1.12 \pm 0.28$ M$_\odot$ from its progenitor white dwarf, and synthesized $\sim 0.44 \pm 0.14$ M$_\odot$ of radioactive $^{56}$Ni.}
   {}

   \keywords{supernovae --
                distance measurement 
               }

   \maketitle
%

\section{Introduction}
\label{introduction}
The Type Ia supernovae (SNe Ia), being high-luminosity standardizable candles, have essential importance in the cosmic distance measurements providing extension of the distance ladder toward higher redshifts. Since at present there is a significant tension between the cosmological parameters, like $H_0$, inferred locally and from the Cosmic Microwave Background (CMB), it is important to further reduce the potential biases in the measured distances, which may help in revealing the cause of the discrepancy.  Local galaxies that hosted SNe Ia and have observable Cepheid populations are especially important in this respect \citep{Riess22}. 

The basis of the standardization process of Type Ia light curves (LCs) is the Phillips relation \citep{Phillips93}, i.e. the empirical correlation between the peak absolute brightness (typically in the $B$-band) and the $\Delta m_\mathrm{15}$ decline rate measured during the first 15 days after the moment of maximum light ($t_\mathrm{0}$). Later, multiple studies tried to link the shape of the LC to the peak luminosity. The most widely used SN Ia LC synthesis and distance estimator codes are the newest versions of SALT \citep{Guy05} and MLCS \citep{Riess98, Jha06}, but other approaches, like BayeSN \citep{Thorp21,Mandel22} and SNooPy \citep{Burns11} have also been published (see Section \ref{lcfitter}).

However, LC fitters and distance estimations still suffer from intrinsic scattering due to spectrophotometric calibration issues and maybe some sort of unknown systematic effects. One way to reduce the sources of uncertainties is using SN siblings, i.e. SNe discovered in the same galaxy. These SNe share the same distance, redshift, and other physical properties of their common host galaxy. Thus, the expected dispersion in their individually estimated distances should be significantly lower than the intrinsic scatter in the Hubble-diagram at the same redshift. Thus, SN siblings may also allow us to test distance measurement methods and may support their further improvements \citep[see e.g. ][]{Burns20, Gallego22, Scolnic20, Hoogendam22, Ward22}. 

Due to the decade-long observations by recent transient discovery programs, the number of galaxies hosting multiple SNe is gradually increasing. The absolute record holder of modern times is NGC 6946, also called the Firework Galaxy, which hosted ten SNe in a century. However, the datasets of early SNe usually suffer either from high uncertainties, data gaps or lack of wavelength coverage, thus, not all of these SNe can be used for high-precision distance estimations. 

We searched for additional supernova siblings in the Open Supernova Catalog \citep[OSC,][]{Guillochon17}. By narrowing our search to the simple Ia category in the OSC, we found 67 galaxies that hosted two or more ``normal'' Type Ia supernovae. The actual number is higher as the OSC has a sub-classification of Ia supernovae that we omitted from our search, and the catalog has not been updated since April 2021. Recent studies by \cite{Burns20} and \cite{Gallego22} have increased the sample further. In the most recent study by \cite{Kelsey23}, 113 galaxies were found to host 236 thermonuclear SNe (including both ``normal'' SNe Ia and other subclasses) in the OSC. 

There are only 6 galaxies in which at least three Type Ia supernova siblings have been discovered. Both M84 (SNe 1957B, 1980I, and 1991bg) and NGC 1316 (SNe 1980N, 1981D, 2006dd, and 2006mr) include two Type Ia supernovae before the CCD era, which increases the importance of NGC 5468 (SNe 1999cp, 2002cr, and 2005P), NGC 5018 (SNe 2017isq, 2002dj, 2021fxy), NGC 3367 (SNe 2018kp, 2003aa, 1986A), and NGC 3147 (SNe 1972A, 1997bq, 2008fv, and 2021hpr). The last one hosted the recently discovered SN 2021hpr, which was the subject of intense follow-up observations by several observatories. Therefore, it became a key object in the distance estimations using Type Ia supernova siblings.

In this paper we present new optical photometric observations of SN~2021hpr, and combine them with published LCs of SNe 1997bq and 2008fv to determine an improved distance to their common host galaxy, NGC~3147. Based on the improved distance, we infer and discuss the physical parameters for SN~2021hpr by building models for its spectra and the bolometric LC. 
The paper is structured as follows: in Section \ref{description}, we introduce the datasets of the three SNe Ia hosted by galaxy NGC 3147, with a special interest in the newly obtained LC of SN 2021hpr published in this paper first.
Methods for the spectral synthesis and LC analysis, as well as the fitting algorithms used for distance estimations, are described in Section \ref{methods}. The results are presented and discussed in Section \ref{results}. Finally, we summarize our conclusion in Section \ref{conclusion}.

\section{Three supernovae of NGC 3147}
\label{description}

 NGC 3147 is a barred spiral galaxy in the Draco constellation, at $\alpha$(2000.0) = $10^\mathrm{h} 16^\mathrm{m} 50^\mathrm{s}$, $\delta$(2000.0) = +73$^\circ$ 24', being in the focus of interest due to its low-luminosity Type II Seyfert active galactic nucleus \citep{Panessa02}. Its heliocentric redshift is $z=0.00934$ \citep{Epinat08}. The historical distance estimations show a wide range between 27.7 Mpc \citep[from Tully-Fisher relation,][]{Bottinelli84} and 55.2 Mpc \citep[based on the Type Ia SN 1972A,][]{Parodi2000}, but the latest pre-SN 2021hpr results narrowed down to 39.3 Mpc \citep[SN 1997bq,][]{Tully13} and 43.7 Mpc \citep[SN 2008fv,][]{Biscardi12}. The most recent Cepheid-based distance is derived by the comprehensive analysis of \cite{Riess22} where the authors estimated 40.1 Mpc $\pm 3.3$ Mpc including the analysis of 27 Cepheid variables of NGC 3147.
 
 The proximity and face-on orientation of NGC 3147 makes it a prominent host for discovering transient events. In the past half-century, this galaxy hosted six SNe, four of which were classified as Type Ia (the other two, SN 2006gi and SN~2021do, were Type Ib/c events). For SN 1972H, only photographic photometry was published \citep{Barbon73}.  Thus, only SNe 1997bq, 2008fv, and the recently discovered 2021hpr can be used in a modern LC analysis. 

The galactic component of the interstellar reddening in the direction of NGC 3147 is $E(B-V)=0.021$ mag \citep{Schlafly11}. The explosions took place at distant sites within NGC 3147, sampling different regions of the galaxy. The host galaxy component of the reddening is taken from previous studies for each object, as listed below.

SN 1997bq was discovered on 50546.0 MJD by \cite{Laurie97}. The SN was located outside the observable spiral arms of NGC 3147 with 60" offset (R $\approx$ 16.0 kpc) southeast from the bulge. $UBVRI$ LCs were obtained at the Fred Lawrence Whipple Observatory of the Harvard-Smithsonian Center for Astrophysics and published by \cite{Jha06}. Due to the outskirt location of SN 1997bq, no significant host galaxy reddening is expected, but the individual LC fit performed with the code BayeSN indicated a total extinction of $A_\mathrm{V} = 0.45$ mag \citep{Ward22}.

SN 2008fv was first detected on 54736.0 MJD by K. Itagaki to 37" north-east from the bulge of NGC 3147 \citep{Nakano08}. \cite{Biscardi12} reported a significant host-galaxy reddening of $E(B-V)_\mathrm{host}=0.22$ with selective extinction coefficients of $R_\mathrm{V}=2.9$ based on \cite{Cardelli89}. The SN peaked at 14.55 mag on $\sim$54749.0 MJD in the $B$-band. Optical and near-infrared (NIR) photometry was obtained by \cite{Biscardi12} and \cite{Tsvetkov10}. However, there is an enormous difference between the two datasets in the $I$-band. The choice of the adopted magnitudes for the LC fitting is explained in Sec. \ref{lcfit_08fv}.

The discovery of SN 2021hpr was reported by \cite{Itagaki21} based on the first observation on 2021-04-02 at 10:46 UTC (59306.4 MJD). Later a pre-discovery detection (59304.92 MJD) was reported by \cite{Tsvetkov21} by 0.3 day before the observation of the Zwicky Transient Factory. The first, yet ambiguous classification \citep{Tomasella21} claimed that the new transient is probably a Type Ia SN due to its Si II $\lambda$6355 feature with an expansion velocity of 21,000 km s$^{-1}$. The strong high-velocity feature (HVF) and the rapid brightening of the object suggested that SN 2021hpr was discovered at its very early phase.

SN 2021hpr was the subject of two previous studies, which published and analyzed their independent follow-up observations: \cite{Zhang22} presented $BVRI$ LCs and optical spectra between -14 and +64 days to $B$-band maximum, while \cite{Ward22} provided $grizy$ photometry between -10 and +40 days and one optical spectrum at -4.2 days. In this study, we publish a new set of optical photometry obtained at two Hungarian observatories. Furthermore, we include the analysis of UV-photometry taken by the Neil Gehrels Swift Observatory Ultraviolet and Optical Telescope (UVOT). Hereafter, we use our new $BVgri$ photometry, supplemented by Swift $UBV$ data,  for the LC analysis of SN~2021hpr.  The previously published LCs mentioned above, are also used for comparison, but not for re-analysis.  
We also model the spectra of SN 2021hpr available from the literature (see Tab. \ref{tab:spectroscopic_data}). 

The host extinction was assumed to be negligible by \cite{Zhang22}, but the LC analysis of \cite{Ward22} suggested $A_{V,host} \simeq 0.20$ mag. Without any more established estimation, we use their mean value,  $A_{V,host}=0.1$ mag, for the rest of the paper.

The main properties of NGC 3147, as well as of the three SNe, are listed in Tab. \ref{tab:parameters}.
  
\subsection{Observations}
\label{observations}
In this paper we present a SN photometric dataset that was obtained with two recently installed 0.8m telescopes in Hungary: one at the Piszkéstető mountain station of Konkoly Observatory, and one at Baja Observatory.
The twin instruments are two 0.8m Ritchy-Chrétien telescopes (hereafter KRC80 and BRC80, respectively), manufactured and deployed by the company AstroSysteme Austria (ASA). The focal length of 5700 mm provides an f/7 light-gathering power. The telescope is equipped with Johnson $BV$ and Sloan $ugriz$  filters and a 2048x2048 back-illuminated FLI PL230 CCD chip with a pixel scale of 0.55". Due to the similarities between the two telescopes, the combined LCs of SN 2021hpr can be considered as a homogeneous dataset. 
 
\begin{figure}
	\includegraphics[width=\columnwidth]{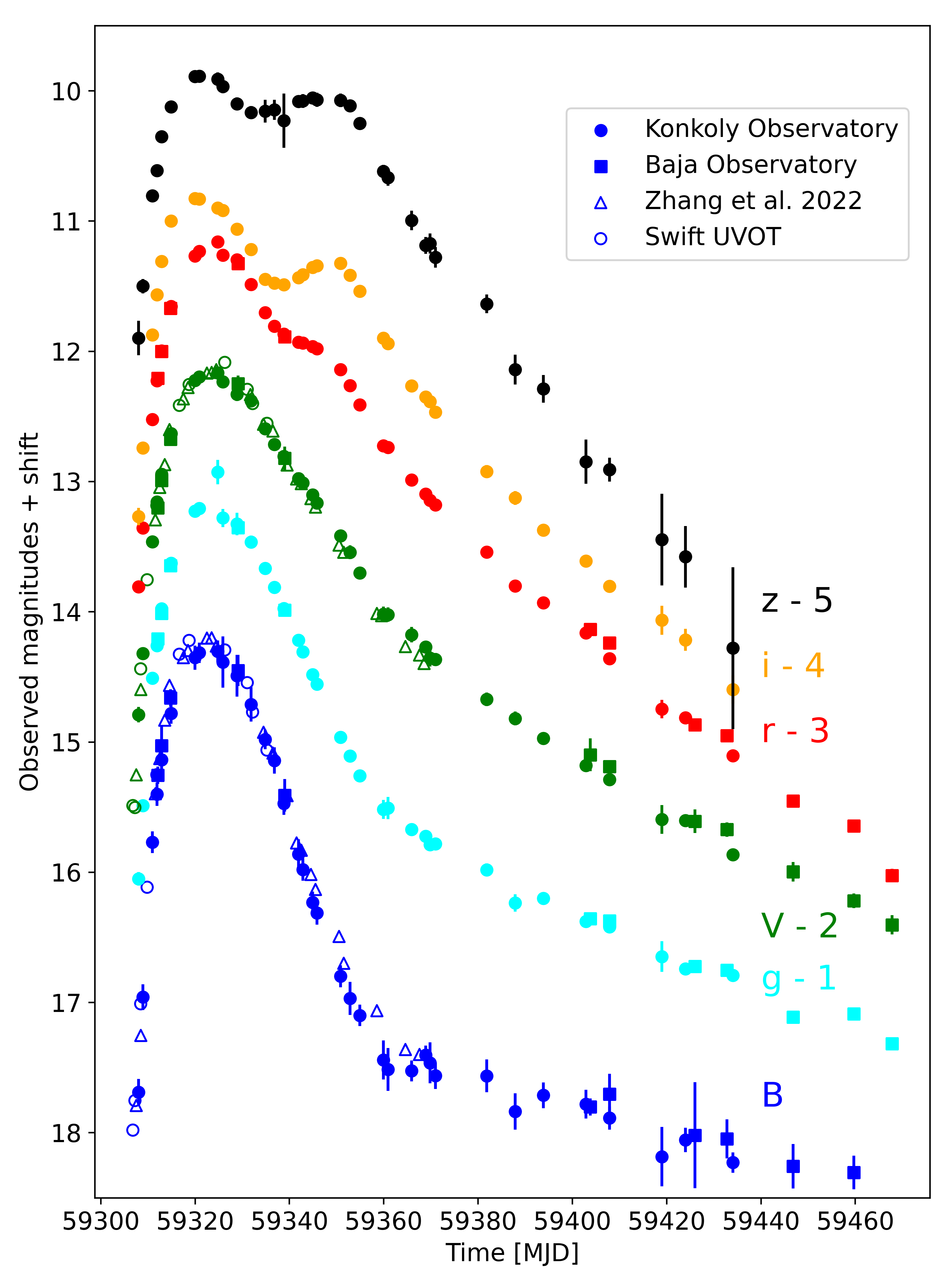}
    \caption{Comparison of $BVgriz$ photometry of SN 2021hpr from different data sources.}
    \label{fig:photometry_21hpr}
\end{figure}

We carried out standard Johnson--Cousins {\it BV} and Sloan {\it griz} CCD observations on 50 nights between April and September 2021 (Fig. \ref{fig:photometry_21hpr}). The achieved photometric accuracy varied between 0.01--0.05 mag depending on the weather conditions. The exposure times were 180 seconds except for the {\it B} filters where we used 300 sec.

All data were processed with standard IRAF\footnote{IRAF is distributed by the National Optical Astronomy Observatories, which are operated by the Association of Universities for Research in Astronomy, Inc., under cooperative agreement with the National Science Foundation.} routines, including bias, dark and flat-field corrections. Then we co-added three images per filter per night aligned with the {\tt wcsxymatch}, {\tt geomap} and {\tt geotran} tasks. We obtained PSF photometry on the co-added frames using the {\tt daophot} package in IRAF, and image subtraction photometry based on other IRAF tasks like {\tt psfmatch} and {\tt linmatch}, respectively. For the image subtraction we applied a template image taken at a sufficiently late phase, when the transient was no longer detectable on our frames. 

For PSF photometry we built an automated pipeline using self-developed C-codes and {\tt bash} shell scripts, and the necessary IRAF tasks are called as system binaries outside the IRAF environment. The IRAF executables are collected into a single parallel processing script using {\tt gnu-parallel} \citep{Tange11}. This ``all inclusive'' method enabled us to reduce the processing time for the $\sim 1$ GB of data per night to a few minutes on a normal PC with 16 CPU cores.

The photometric calibration was carried out using
stars from Data Release 1 of Pan-STARRS1 (PS1 DR1) \footnote{https://catalogs.mast.stsci.edu/panstarrs/}. The selection of the photometry reference stars and the
calibration procedures are as follows. First, sources within a 5 arcmin radius around the SN with $r$-band brightness between 15 and 17 mag (to avoid saturation, \citep{Magnier13}) were downloaded from the PS1 catalog. Next, non-stellar sources were filtered out based on the criterion $i_\mathrm{PSFmag} - i_\mathrm{Kronmag}$ < 0.05 for stars \footnote{https://outerspace.stsci.edu/display/PANSTARRS/}. 
In order to get reference magnitudes for our Johnson $B$- and $V$-band frames, the PS1 magnitudes were transformed into the Johnson BVRI system based on equations and coefficients found in \cite{Tonry12}. Finally, the instrumental magnitudes were transformed into standard $BVgriz$ magnitudes by applying a linear color term (using $g-i$) and wavelength-dependent zero points. Since the reference stars fell within a few arcminutes around the target, no atmospheric extinction correction was necessary. 
S-corrections were not applied.


The obtained $BVgriz$ LCs are plotted in Fig. \ref{fig:photometry_21hpr}. Direct comparison with $B$- and $V$-band LCs of \cite{Zhang22} further confirm that our data are free of systematic errors.

\begin{figure}
	\includegraphics[width=\columnwidth]{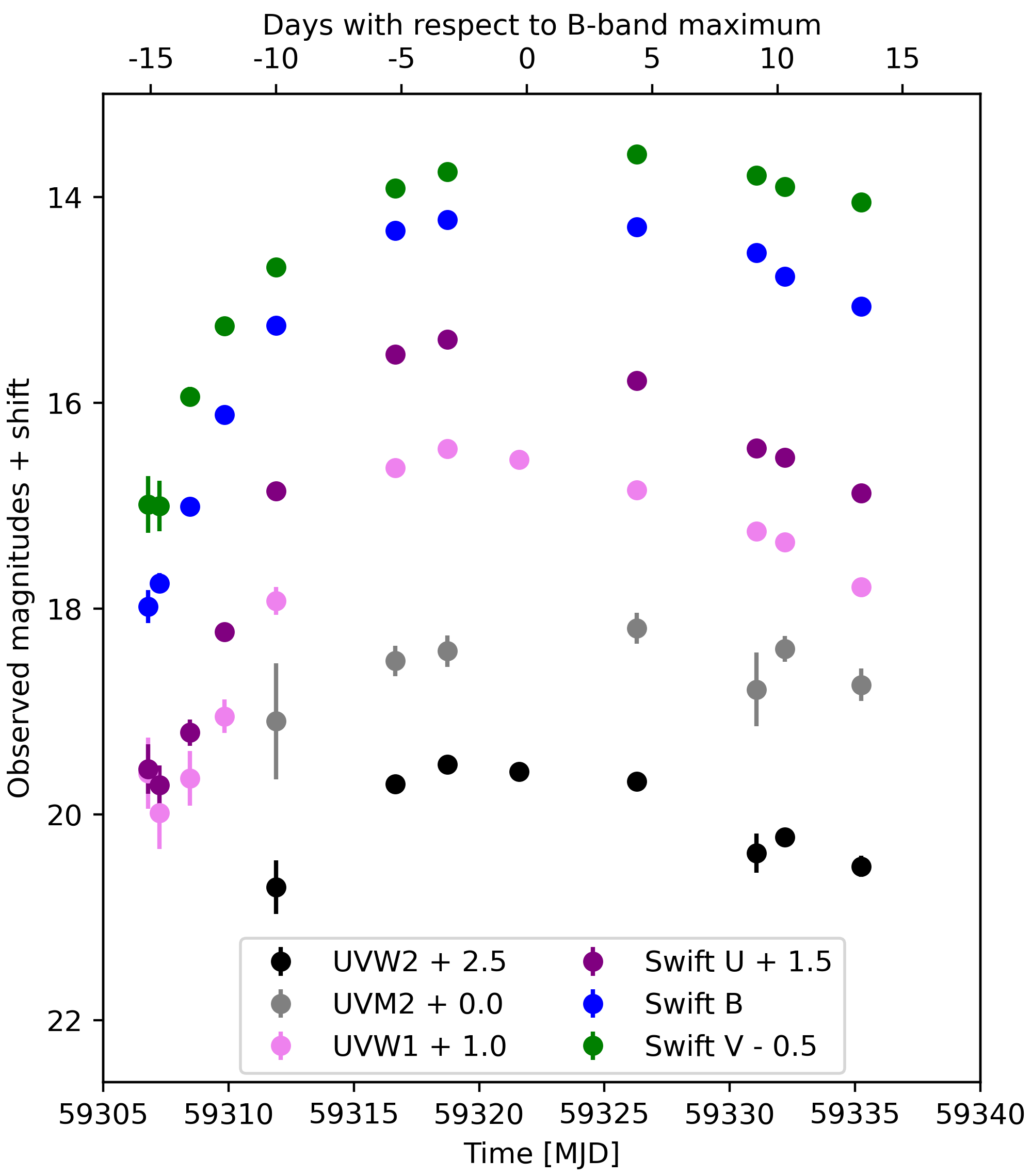}
    \caption{Swift photometry of SN 2021hpr.}
    \label{fig:swift_21hpr}
\end{figure}

In the case of SN~2021hpr the ground-based optical observations were supplemented by the available archival data of the Neil Gehrels Swift Observatory \citep[{\it Swift},][]{Gehrels04, Burrows05} taken with the Ultraviolet-Optical Telescope \citep[\textit{UVOT},][]{Roming05} in April 2021 (see in Fig. \ref{fig:swift_21hpr}). Data were collected in six filters from optical to ultraviolet wavelengths (\textit{u, b, v, uvw1, uvm2, uvw2}). The SN was detectable as a point source on the images, although it was located in a complex galactic environment. In order to model its background flux, we applied five different background regions distributed around the SN, and determined the background as the average of the flux values taken from each region. The \textit{Swift}/UVOT data were processed using the HEAsoft software package. We summed the individual frames using the \texttt{uvotimsum} task and carried out aperture photometry on the summed images using the \texttt{uvotsource} task.

Two spectra used in this study (see Tab. \ref{tab:spectroscopic_data}) were published by \cite{Zhang22}, and another was obtained at Smolecin Observatory (L25). All spectra are available at WISeREP online supernova database \citep{Yaron12}.

\section{Methods}
\label{methods}

In this section, we briefly introduce the modeling codes adopted for the photometric and spectroscopic analyses.

  \begin{table}
	\centering
	\caption{Main parameters of the three SNe Ia in NGC 3147.}
	\label{tab:supernovae}
	\begin{tabular}{l|cc} 
		\hline
            \multicolumn{2}{c}{\textbf{NGC 3147}} & Ref. \\
            \hline
            heliocentric redshift & \multicolumn{1}{c}{0.00934} & 1 \\
            $E(B - V)_\mathrm{MW}$ & \multicolumn{1}{c}{0.021 mag} & 2 \\
            \hline
             \hline
		\multicolumn{3}{c}{SN 1997bq}\\
		\hline
		RA & 	10h 17m 04s & 3\\
		DEC & 	+73$^\circ$ 23' 03" & 3\\
            T$_\mathrm{max}(B)$ & 50558.0 MJD & 4 \\
            B$_\mathrm{max}$ & 14.57 mag & 4 \\
            $\Delta m_\mathrm{15}(B)$ & 1.01 mag & 4 \\
		$E(B - V)_\mathrm{host}$ & 0.11 mag & 5\\
		\hline
             \hline
		\multicolumn{3}{c}{SN 2008fv}\\
            \hline
		RA & 	10h 16m 57s.28 & 6\\
		DEC & 	+73$^\circ$ 24' 36" & 6\\
            T$_\mathrm{max}(B)$ & 54749.3 MJD & 7 \\
            B$_\mathrm{max}$ & 14.55 mag & 7 \\
            $\Delta m_\mathrm{15}(B)$ & 0.94 mag & 7 \\
		$E(B - V)_\mathrm{host}$ & 0.22 mag & 7\\
            \hline
             \hline
		\multicolumn{3}{c}{SN 2021hpr}\\
            \hline
		RA & 	10h 16m 38s.68 & 8\\
		DEC & 	+73$^\circ$ 24'01 00".80 & 8\\
            T$_\mathrm{max}(B)$ & 59321.9 MJD & 9 \\
            B$_\mathrm{max}$ & 14.017 mag & 9 \\
            $\Delta m_\mathrm{15}(B)$ & 0.949 mag & 9 \\
		$E(B - V)_\mathrm{host}$ & 0.00/0.07 mag  & 9,5\\
            \hline
\multicolumn{3}{l}{\smallskip}\\
        \multicolumn{3}{l}{1 - \cite{Epinat08}; 2 - \cite{Schlafly11};} \\ 
         \multicolumn{3}{l}{3  \cite{Laurie97}; 4 - \cite{Jha06};} \\
          \multicolumn{3}{l}{5 - \cite{Ward22}; 6  \cite{Nakano08};} \\
           \multicolumn{3}{l}{7 - \cite{Biscardi12}; 8 - \cite{Itagaki21};}\\ 
           \multicolumn{3}{l}{9 - \cite{Zhang22}}
    \label{tab:parameters}
\end{tabular}
\end{table}

\subsection{The radiative transfer code TARDIS}
\label{sec:tardis}

Our approach for the fitting of the spectral series of SN 2021hpr is performed with the one-dimensional radiative transfer code {\tt TARDIS} \citep{Kerzendorf14}. {\tt TARDIS} calculates synthetic spectra on a wide wavelength range in exchange for a low computational cost, providing an ideal tool for fitting both the continuum and spectral features of homologously expanding ejecta. 

The main assumption of the code is a sharp photosphere emitting blackbody radiation modeled via indivisible energy-packets representing bundles of photons \citep[for further description see ][]{Abbott85,Lucy93, Lucy99, Lucy02, Lucy03}. The model atmosphere is divided into radial layers with densities and chemical abundances defined by the user. The algorithm follows the propagation of the packets, while calculating the wavelength- and direction-changing light-matter interactions based on a Monte Carlo scheme in every shell. The final output spectrum is built up by summarizing the photon packets escaping from the model atmosphere.

The approach of {\tt TARDIS} offers several improvements from the simple LTE assumptions. We followed the same settings as in most of the studies that used {\tt TARDIS} for spectral fitting \citep[see e.g.][]{Magee16, Boyle17, Barna18}. As an example, the ionization state of the material used here are estimated following an approximate non-LTE (NLTE) mode, the so-called nebular approximation, which significantly deviates from the LTE method by accounting for a fraction of recombinations returning directly to the ground state \citep{Mazzali93}. The excitation state is calculated according to the dilute-LTE approximation which is also not purely thermal. The summary of {\tt TARDIS} numerical parameters and modes adopted in this study are listed in Tab. \ref{tab:tardis}. The limitation of the simulation background, as well as the detailed description of the NLTE methods, are presented in the original {\tt TARDIS} paper \citep{Kerzendorf14}.

  \begin{table}
	\centering
	\caption{{\tt TARDIS} fitting parameters}
	\label{tab:tardis}
	\begin{tabular}{lc} 
		\hline
		  Setting & Approximation \\
		\hline
            Radiation mode & dilute-blackbody \\
		Ionization mode & nebular\\
		  Excitation mode & dilute-LTE\\
		Line interaction mode & macroatom \\
            \hline
            Parameter & Best-fit value \\
            \hline
            Time of explosion & 59304.0 MJD\\
            Core density ($\rho_0$) & 4.7 $g\,cm^{-3}$\\
            Density slope ($v_\mathrm{0}$) & 2750 $km\,s^{-1}$\\
            
		\hline
	\end{tabular}
\end{table}

\subsection{Light curve fitter codes for SNe Ia}
\label{lcfitter}

We applied the recently released version of Spectral Adaptive Lightcurve Template \citep[SALT3,][]{Kenworthy21} for estimating the distance to SN 2021hpr and its two siblings. We also utilized the earlier version, SALT2.4 \citep{Betoule14}, as well as an independent LC fitter, MLCS2k2 \citep{Jha07} for checking the consistency of the inferred distance moduli with those from earlier calibrations.   

MLCS2k2 is the improved version of the original Multi-Color Light Curve Shape (MLCS) code introduced by the High-z Supernova Search Team \citep{Riess96}. The LCs obtained in Johnson-Cousins $UBVRI$ filters are fitted with two tabulated functions ($P_\lambda$ and $Q_\lambda$) trained on a carefully selected sample of 133 SNe \citep{Jha07} and with the $\Delta$ parameter that linked the shape of the LC with the peak absolute brightness in the $V$-band. The code fits the observed LC $m_\lambda (t)$ with the following function
\begin{equation}
m_\lambda (t) = M_\lambda (t) + \mu_0 + \zeta_\lambda \big( \alpha_\lambda + \beta_\lambda / R_V \big) \cdot A_V + P_\lambda(t) \cdot \Delta + Q_\lambda(t) \cdot \Delta^2, 
\end{equation}
where $t$ is the rest-frame time (corrected for time dilation) elapsed from the moment of $B$-band maximum, $M_\lambda (t)$ is the LC of the fiducial SN Ia in absolute magnitudes, and $\mu_0$ is the true (extinction-free) distance modulus. MLCS2k2 also takes into account the effect of interstellar reddening with $A_V$ extinction in the $V$-band. $\alpha$, $\beta$ describe the interstellar reddening law as a function of wavelength, while $\zeta$ takes into account the temporal variation of the reddening correction due to the spectral evolution of the SN. 

SALT2 was introduced by the SuperNova Legacy Survey Team \citep{Guy07}. Unlike MLCS2k2, this code models the entire spectral energy distribution (SED) as a function of time. To do this, a combination of multiple colour-dependent vectors is trained on a large sample of thoroughly chosen SNe. The components of the model function are
\begin{equation}
F(\lambda,t) = x_\mathrm{0} \cdot [M_\mathrm{0}(\lambda,t) + x_\mathrm{1} \cdot M_\mathrm{1}(\lambda,t)] \cdot \exp(c \cdot CL(\lambda)),
\end{equation}
where $M_0$, $M_1$ and CL are the trained vectors of SALT, while $x_0$, $x_1$ and $c$ are fitting parameters, representing the normalization, the stretch and the colour of the SED, respectively.


SALT3 is a recent improvement to SALT2.4 developed by \cite{Kenworthy21}. SALT3 was trained on more than one thousand SN Ia spectra, an order of magnitude larger sample than the previous version of the code, providing lower uncertainties and fewer systematics compared to the previous versions.

%

The SALT model does not contain the distance as a direct fitting parameter. Instead, it is inferred from the fitting parameters of the SALT3 code by the following formula \citep{Tripp98}:

\begin{equation}
  \mu = -2.5 \log_{10}(x_0) + \alpha x_1 - \beta c - M_0
\end{equation}

We adopt the recent calibration of \cite{Pierel22} for $\alpha$ and $\beta$ parameters as $\alpha = 0.133 \pm 0.003$ and $\beta = 2.846 \pm 0.017$, respectively. 

\section{Results}
\label{results}

\subsection{Distance measurements}
\label{distance}

Recently, a comprehensive analysis of the distance of a multiple-SNe host galaxy was performed by \cite{Ward22}. The authors presented a new version of the BayeSN model \citep{Thorp21, Mandel22} fitting the optical-NIR regime (in $grizy$ bands) of SN 2021hpr, but also included the photometry of SNe 1997bq and 2008fv, both individually and simultaneously. They inferred the mean value of the distance moduli as $\mu = 33.14$ mag with a standard deviation of 0.01, smaller than the intrinsic scattering of SNe Ia. The common $\mu$-fit to the data of the three SNe resulted in $\mu = 33.13 \pm 0.08$ mag. 

\cite{Ward22} also compared BayeSN to the widely used LC-fitter code SNooPy. The latter code provided a higher standard deviation of $\sigma \sim 0.1$ mag, but the mean value of the individual three distance moduli ($\mu = 33.18$ mag) was close to that of the BayeSN code.

\begin{figure}
	\includegraphics[width=\columnwidth]{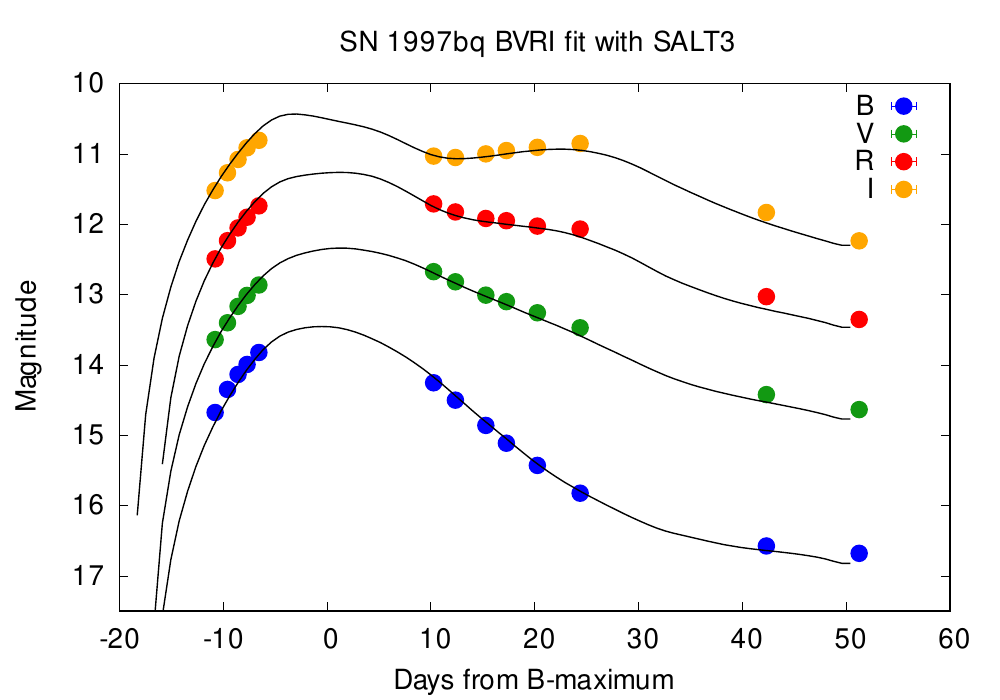}
    \caption{SALT3 LC model fitting of the photometry of SN 1997bq}
    \label{fig:salt3_97bq}
\end{figure}

\begin{figure}
	\includegraphics[width=\columnwidth]{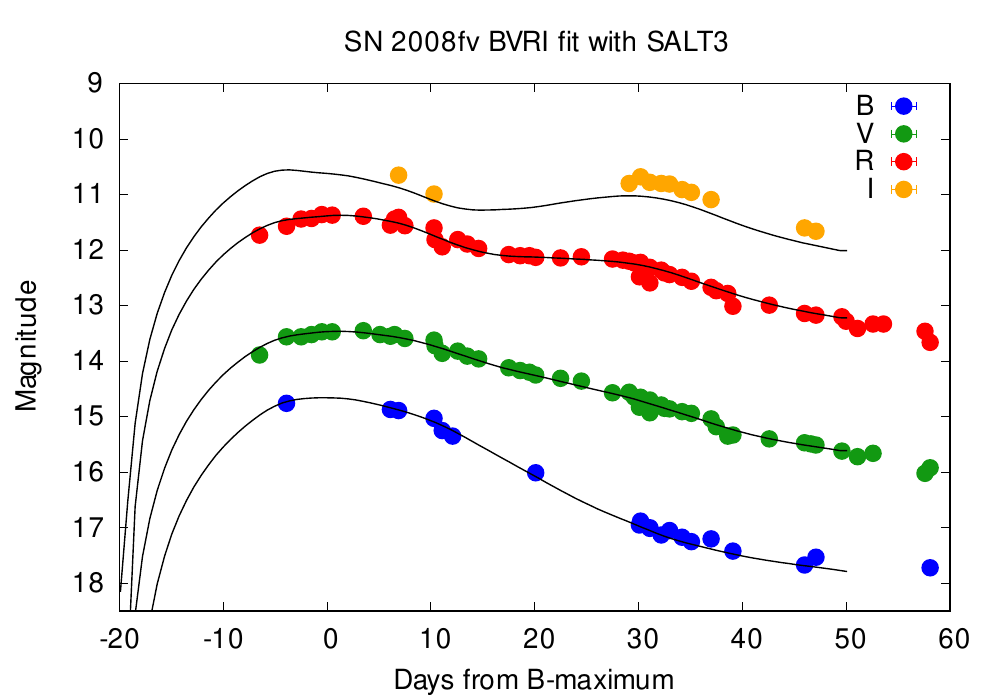}
    \caption{SALT3 LC model fitting of the photometry of SN 2008fv.}
    \label{fig:salt3_08fv}
\end{figure}

\begin{figure}
	\includegraphics[width=\columnwidth]{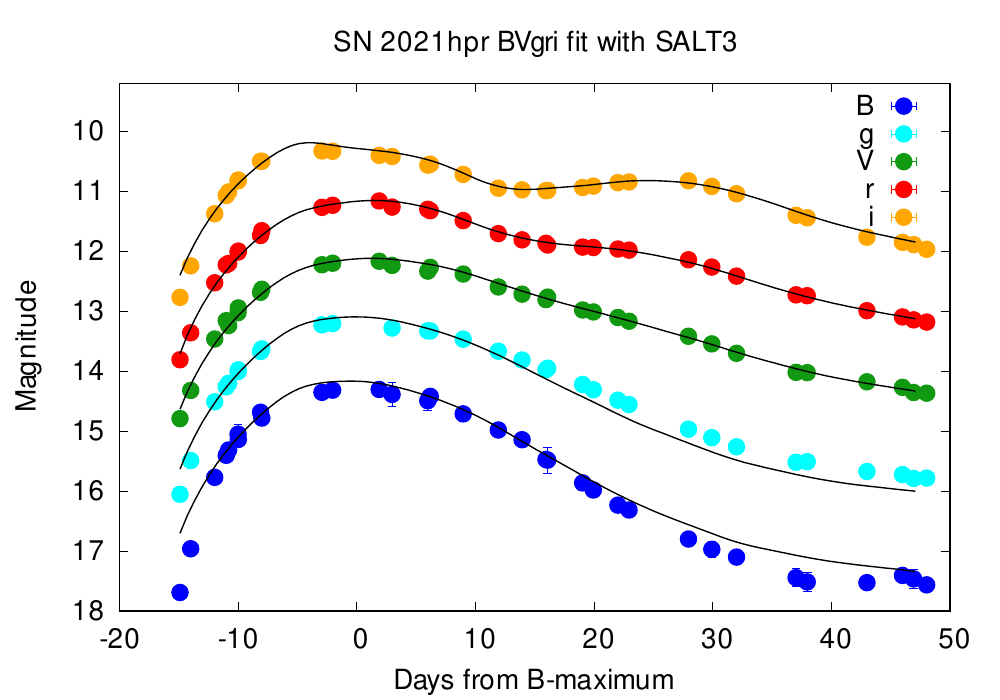}
    \caption{SALT3 LC model fitting of the photometry of SN 2021hpr.}
    \label{fig:salt3_21hpr}
\end{figure}

In the present paper, we apply SALT3 for a similar analysis as \cite{Ward22} and use SALT2.4 and MLCS2k2 for cross-comparing the inferred distance moduli. All LC fittings are computed separately on the individual datasets of SNe 1997, 2008fv, and 2021hpr. While SALT3 can be fed by a collection of various filters and magnitude systems, MLCS2k2 was trained for Johnson-Cousins $UBVRI$ filters and Vega magnitudes. In the case of SN 2021hpr, where the object was not observed in Johnson-Cousins $R$ and $I$ filters, only $B$ and $V$ data were used during the fitting with MLCS2k2. SALT3 fits are shown in Figs. \ref{fig:salt3_97bq} - \ref{fig:salt3_21hpr}, while MLCS2k2 and SALT2 fits can be seen in Figs. \ref{fig:mlcs_97bq} - \ref{fig:salt2_21hpr5}.


Due to the low ($z < 0.01$) redshift of the host galaxy, K-correction for transforming observed LCs to rest-frame bands is estimated in the order of 0.01 mag, which is lower than the random observational uncertainties of the individual data, thus, K-corrections were neglected. To avoid any discrepancy due to the different H$_\mathrm{0,ref}$ values that were assumed to tie LC fitting codes to the distance scale, all distances are transformed to a common value of H$_0 = 73$ km s$^{-1}$ \citep{Riess22}:

\begin{equation}
\mu (H_0) = \mu_\mathrm{fit} - 5\, \log(H_0 / H_\mathrm{0,ref})
\label{eq:h0}
\end{equation}

where $\mu_\mathrm{fit}$ is the distance modulus inferred directly by the LC fitter. 
Despite that all three LC fitter codes have been trained for the U-band wavelengths, we refrain from using the U-band (both Johnson and Swift) magnitudes. The near-UV diversity of SNe Ia is not fully covered, because of the limited sample of observations in the U-band. Thus, the training of the LC fitters cannot be sufficient for this spectral regime and increase the uncertainty of the distance estimation.

The individual distance estimations based on the three SNe and carried out with the three LC fitters are summarized in Tab. \ref{tab:distance_moduli}.

\subsubsection{SN 1997bq}
The $BVRI$ LCs of SN 1997bq are deficient around the peak, which makes constraining the date of maximum light more difficult. The polynomial fit of $B$-band LC provided $T_{max}(B)=50557.5$ MJD which is consistent with the result of SALT when the time of maximum is a fitting parameter ($T_{max}(B)=50558.0$ MJD). Adopting the latter date for the MLCS2k2, the inferred host galaxy extinction is $A_\mathrm{host}=0.60$ mag, and the constrained distance modulus, $\mu=33.00$ mag, is close to the Cepheid-distance \citep[$\mu_{Ceph} = 33.014 \pm 0.165$ mag, ][]{Riess22}.

The SALT distances perfectly coincide with each other and also support our final conclusion about the distance of NGC 3145 (see below). Moreover, it is in between the previous estimates published by \cite{Tully13} and \cite{Jha07} (32.97 and 33.16 mag, respectively, assuming $H_0 = 73$ km\,s$^{-1}$, see Eq. \ref{eq:h0}) based on the same photometric data of SN 1997bq.

\subsubsection{SN 2008fv}
\label{lcfit_08fv}
For SN 2008fv the publicly available $I$-band observations of \cite{Biscardi12} are significantly brighter than expected, and they exceed the published $I$-band magnitudes of \cite{Tsvetkov10} by $\sim$0.8 mag. \cite{Ward22} investigated the colours of SN 2008fv and concluded that the $I$-band data of \cite{Biscardi12} are probably inaccurate, thus, they were discarded from the fitting process. The \cite{Biscardi12} photometry in other filters was also discredited by \cite{Ward22}, but the estimated discrepancy is in the order of 0.1 mag. In this study, we aimed to use most of the available data, thus, we adopted the complete $BVR$ photometry of \cite{Biscardi12} together with post-maximum observations of \cite{Tsvetkov10} in $BVRI$. 

The MLCS2k2 code provides $\mu_\mathrm{08fv} = 32.97$ mag, which is in good agreement with the Cepheid-based distance modulus of the host galaxy \citep{Riess22}. The total $A_\mathrm{V}$ is constrained as 0.96 mag, significantly higher than that estimated by \cite{Ward22} with the BayeSN code. At the same time, the nearly identical models of SALT2.4 and SALT3 fail to fit the $I$-band fluxes and infer significantly higher distance moduli of $\mu \sim 33.33 \pm 0.09$ and $\mu \sim 33.35 \pm 0.05$ mag, respectively.

\subsubsection{SN 2021hpr}
\label{lcfit_21hpr}
SN 2021hpr has the most densely covered light curves among the three SNe extending from $-15$ to $+100$ days in $BVRI$ and $griz$ bands. However, the $z$-band LC suffers from higher uncertainties due to inferior sky conditions during the observations, thus, it was omitted from the fitting.

The MLCS2k2 fit is made using only the $BV$ LCs from the (B)RC80 observations to keep the homogeneity of the dataset and avoid any systematic errors. However, the result ($\mu = 32.85$ mag) greatly differs from any other distance modulus of this study, which underlines the importance of having a photometric dataset covering wide spectral range. At the same time, the SALT codes provide good fits to all of the LCs in all bands around the maximum light, resulting in similar distance moduli within a $1 \sigma$ agreement. As a further validation, we also include the $BVRI$ photometry of \cite{Zhang22} to the homogeneous $BVgriz$ dataset for an additional modeling with SALT3. The extra LCs and data points barely change the parameters of the fits (Fig. \ref{fig:salt3_21hpr6}), including the inferred distance modulus of $\mu = 33.15$ mag.

\begin{figure*}
	\centering
	\includegraphics[width=15cm]{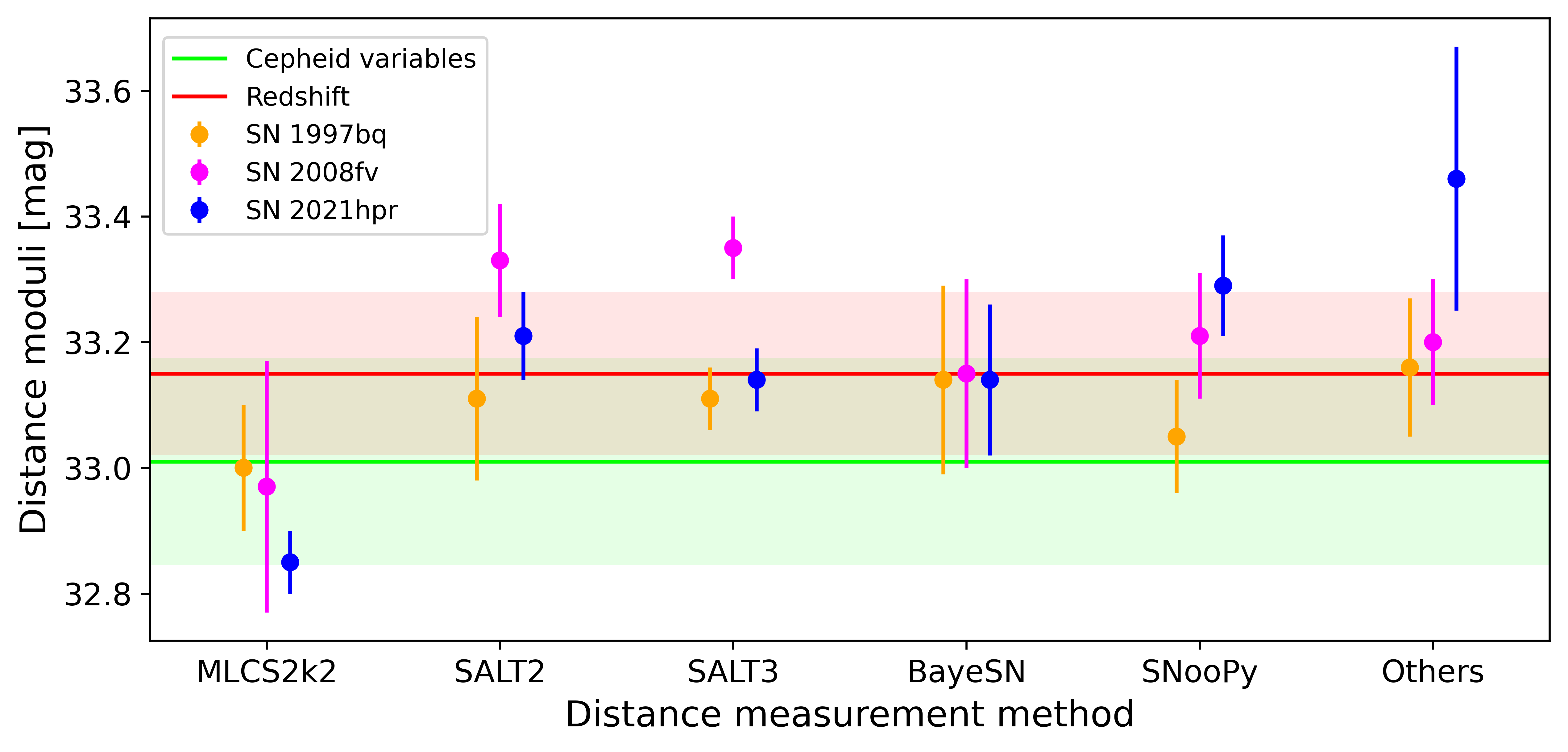}
    \caption{Summary of the individual distance measurements involving the three SNe of NGC 3147. The MLCS2k2, the SALT2 and the SALT3 modeling is published in this study; the distance estimates conducted with BayeSN and SNooPy were published by \cite{Ward22}, while the 'Others' refers to the work of \cite{Jha06}, \cite{Biscardi12} and \cite{Zhang22}. The most recent distance estimations based on Cepheid-variables and cosmological redshift are also shown with horizontal lines.}
    \label{fig:dist_moduli}
\end{figure*}

As a conclusion, we propose the distance modulus $\mu = 33.14 \pm 0.05$ mag of SN 2021hpr estimated by SALT3 for the distance of NGC 3147. This distance shows a good match with the mean value of (\textit all) the distance moduli (including SNe 1997bq and 2008fv) estimated in this study ($\mu = 33.12 \pm 0.10$ mag), and also with that of \cite{Ward22} ($\mu = 33.14 \pm 0.12$ mag). Moreover, the inferred distance is consistent with the result of the Cepheid-based distance \citep[$\mu = 33.01 \pm 0.165$ mag][]{Riess22}. The summary of the distance estimation of NGC 3147 published in the literature and inferred in this study can be found in Fig. \ref{fig:dist_moduli} and in Tab. \ref{tab:distance_summary}.

 \begin{table}
	\centering
	\caption{Comparison of the distance moduli estimated with different LC fitter codes for the three studied SNe of NGC 3147.}
	\label{tab:distance_moduli}
	\begin{tabular}{lccc} 
		\hline
		Code & SN 1997bq & SN 2008fv & SN 2021hpr\\
		\hline
            \textbf{SALT3} & \textbf{33.11 $\pm$0.05} & \textbf{33.35 $\pm$ 0.05} & \textbf{33.14 $\pm$ 0.05}\\
		MLCS2k2 & 33.00 $\pm$0.10 & 32.97 $\pm0.20$ & 32.85 $\pm$ 0.05\\
		SALT2 & 33.11 $\pm$0.13 & 33.33 $\pm$ 0.09 & 33.21 $\pm$ 0.07\\
	
		\hline
	\end{tabular}
\end{table}

  \begin{table*}
	\centering
	\caption{Summary of the most recent distances estimated with approaches. All distance moduli are scaled according to $H_0 =73$ km\,s$^{-1}$\,Mpc$^{-1}$. }
	\label{tab:distance_summary}
	\begin{tabular}{l|c|c} 
		\hline
            Method & $\mu_\mathrm{NGC3147}$ & Reference \\
            \hline
            Cepheid-variables & 33.01 $\pm$ 0.165 & \cite{Riess22} \\
            Redshift & 33.15 $\pm$ 0.15 & \cite{Mould00} \\
            Tully-Fisher relation & 33.12 $\pm$ 0.80 & \cite{Tully88} \\
            \hline
            SN 1997bq (MLCS2k2) & 33.16 $\pm$ 0.11 & \cite{Jha06} \\
            SN 2008fv (Phillips-relation) & 33.20 $\pm$ 0.10 & \cite{Biscardi12} \\
            \hline
            SN 2021hpr (Phillips-relation) & 33.46 $\pm 0.21$ & \cite{Zhang22} \\
            SN 2021hpr (BayeSN) & 33.14 $\pm$ 0.12 & \cite{Ward22} \\
            {\bf SN 2021hpr (SALT3)} & {\bf 33.14 $\pm$ 0.05} & {\bf This study} \\
            \hline
            Common $\mu$-fit of SNe (BayeSN) & 33.13 $\pm$ 0.08 & \cite{Ward22} \\
            Average of LC fits of SNe (MLCS2k2, SALT2, SALT3) & 33.12 $\pm$ 0.16 & This study \\
            \hline

\end{tabular}
\end{table*}

\subsection{The physical properties of SN 2021hpr}

\subsubsection{Rise time}
\label{exp_date}

To constrain the moment of the explosion of SN 2021hpr, we adopted the assumption of the expanding fireball model. According to this, the emerging pre-maximum flux increases as a power-law function of time \citep{Arnett82,Nugent11}:
\begin{equation}
F = a \cdot \left(T - T_\mathrm{first}\right)^{n}
\label{eq:fireball}
\end{equation}
Here, $T_\mathrm{first}$ is the time of first light, which is not equivalent to the time of the
explosion ($T_\mathrm{exp}$), as it refers to the moment when the first photons emerge,
while the latter refers to the actual moment when the explosion starts. 
The intermediate ‘dark phase’ may last for a few hours to days \citep{Piro13,Firth15}.

In theory, the power-law function with $n=2$ is valid only for the bolometric flux. However, studies pointed out that it is still more-or-less valid for quasi-monochromatic fluxes in the optical bands, but in those cases the value of the exponent may differ from the textbook example, varying between $n=2.2$ \citep{Ganeshalingam11} and $n=2.44$ \citep{Firth15} for normal SNe Ia.

We fit the early magnitudes of the most densely sampled LCs of SN~2021hpr, i.e. the B and V bands, simultaneously with the function Eq. \ref{eq:fireball}, but using the same $T_\mathrm{first}$ for both light curves. At first, we fix the exponent as $n=2$, which corresponds to the classical fireball model. The resulting moment of first light is $59305.4 \pm 1.5$ MJD, which is slightly late in the aspect of the first detection a day later.

\begin{figure}
	\includegraphics[width=\columnwidth]{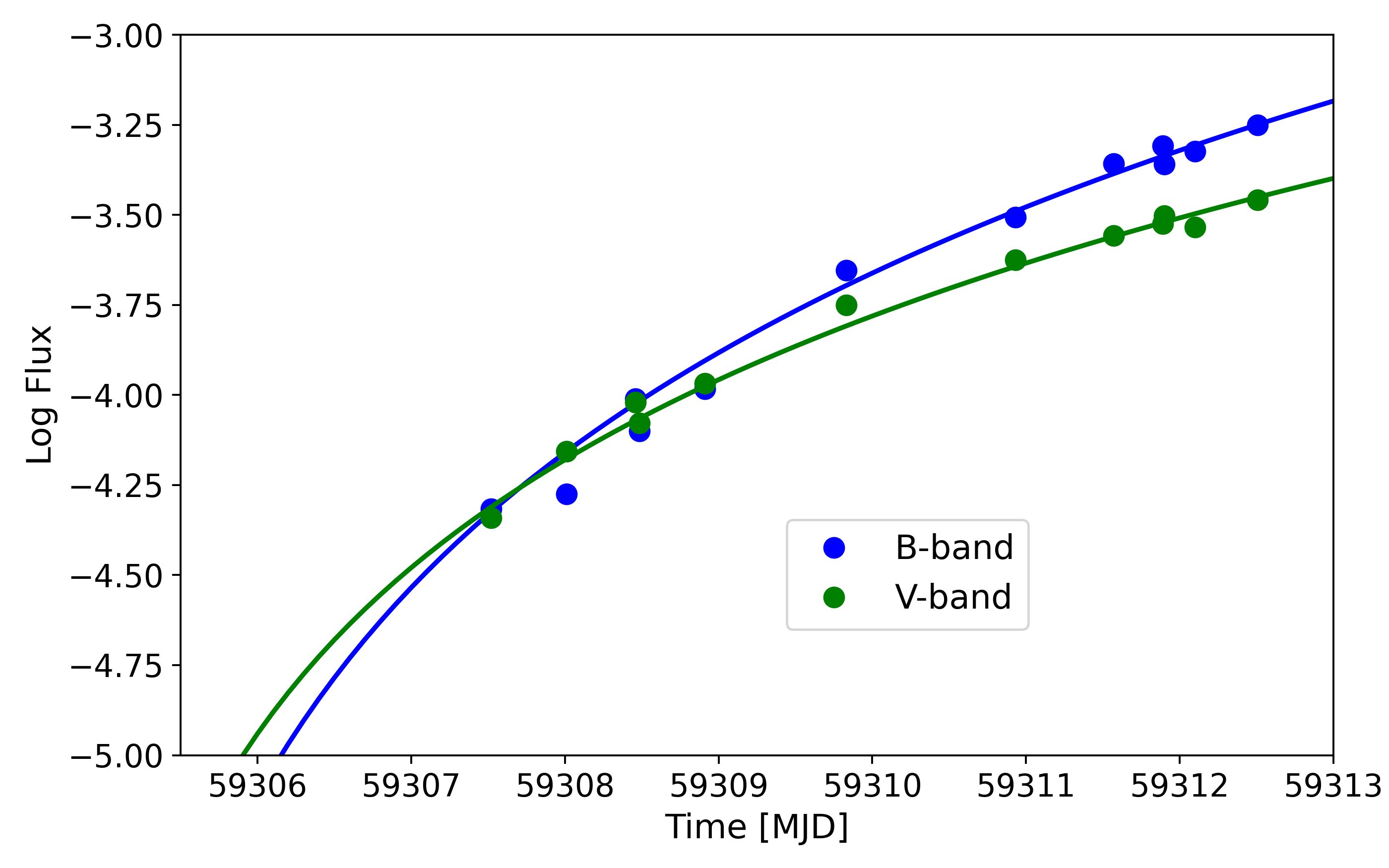}
    \caption{The fit of $BV$ LCs with the function \ref{eq:fireball}, allowing the exponent to vary between 2.0 < $n$ < 2.5. A shared time of first light ($T_\mathrm{first}$) is assumed for both LCs.}
    \label{fig:fireball_21hpr}
\end{figure}

Next, we allow $n$ to vary between 2.0 and 2.5 for each LCs (Fig. \ref{fig:fireball_21hpr}). The inferred date of $T_\mathrm{first} = 59304.6 \pm1.6$ MJD is a more realistic time for the explosion, considering the early discovery 1.9 days later. The LC in the $B$-band peaks at 59323.0 MJD as it is constrained by polynomial fit. As a key parameter, this date is used as input for the MLCS2k2 and SALT2 fits in the following. The estimated 18.4 days rise time is in good agreement with the average of normal SNe Ia \citep[18.98 $\pm$ 0.54 days,][]{Firth15}.

\subsubsection{Abundance tomography}
\label{tardis}

We carried out modeling the spectroscopic evolution of SN~2021hpr with the radiative transfer code {\tt TARDIS} \citep{Kerzendorf14}. To synthesize the spectral luminosities, we fixed the distance of SN 2021hpr to 42.5 Mpc corresponding the distance modulus carried out with SALT3 (see in Sec. \ref{lcfit_21hpr}). 
For de-reddening, a total of $A_\mathrm{V}=0.17$ mag was adopted as the average value of the total extinction assumed by \cite{Zhang22} and \cite{Ward22}.

\begin{figure}
	\includegraphics[width=\columnwidth]{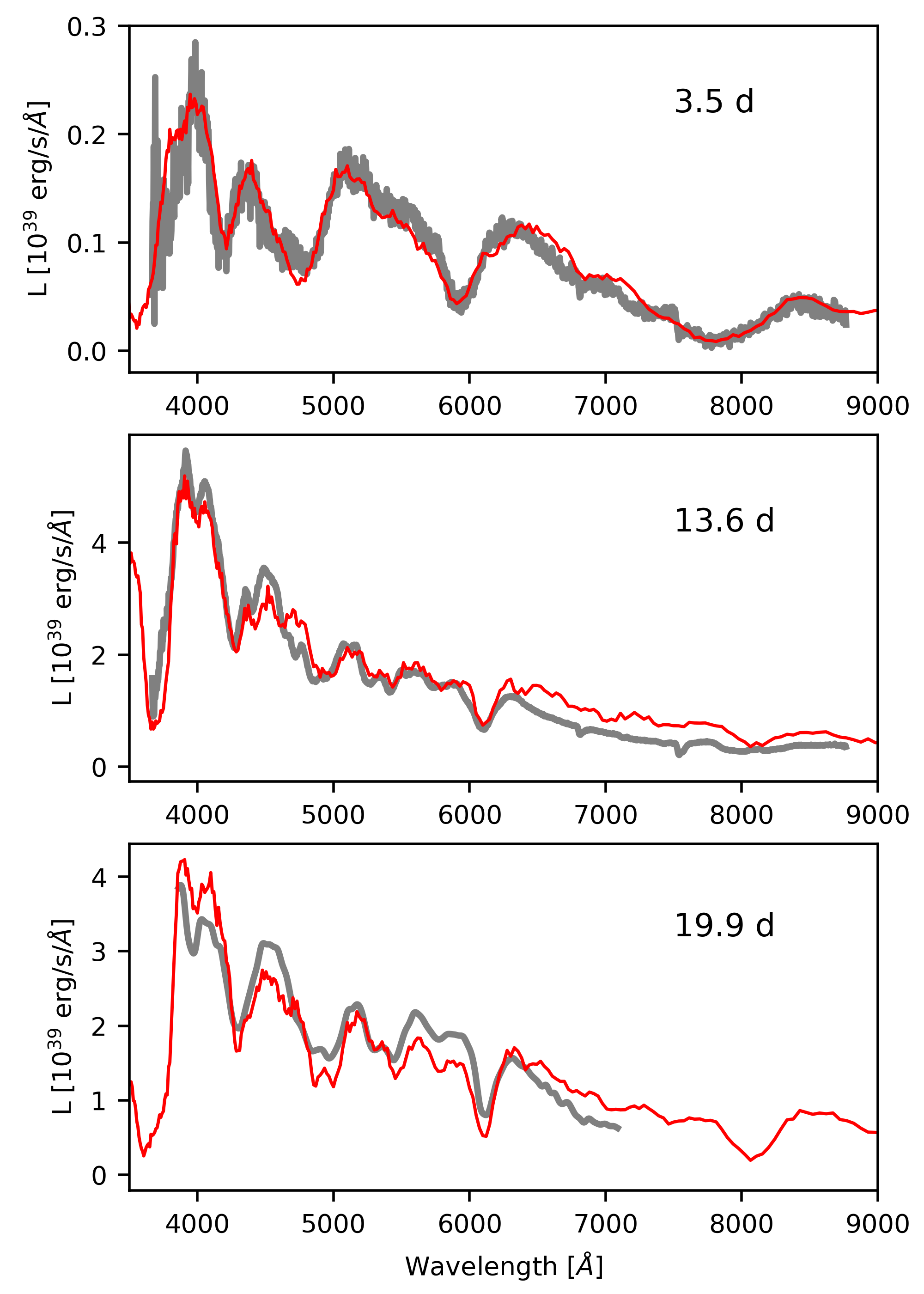}
    \caption{The {\tt TARDIS} fits (red lines) of the three spectral epochs (grey), time since explosion ($t_\mathrm{exp}$) is indicated in upper right corner.}
    \label{fig:tardis_fits_21hpr}
\end{figure}

Three of the earliest spectra of SN 2021hpr have been subject to fit (see in Tab. \ref{tab:spectroscopic_data}). The best-fit synthetic spectra are presented in Fig. \ref{fig:tardis_fits_21hpr}. The key parameters of the spectral synthesis are the total bolometric luminosity ($L$), the photospheric velocity ($v_\mathrm{phot}$), and the time since the explosion ($t_\mathrm{exp}$, derived from $T_\mathrm{exp}$). We fixed the density function of our input to the well-known W7 model \citep{Nomoto84} to reduce the number of free parameters. The latter simplification results in a discrepancy, if we assume the constrained $T_\mathrm{first}$ (see in Sec. \ref{exp_date}) directly as $T_\mathrm{exp}$, because the diluted density structure causes a too-dense and too-hot model ejecta, especially for the first epoch (59307.5 MJD). To compensate for this, we choose to fit $T_\mathrm{exp}$ in our abundance tomography within the range of one day before $T_\mathrm{first}$, taking into account an approximate dark phase. The best value is characterized as $T_\mathrm{exp} = 59304.0$ MJD. 

The spectral tomography taken at the earliest epoch samples the outermost layers, which is located  above 18\,000 km s$^{-1}$ based on the $v_\mathrm{phot}$ at $t_\mathrm{exp} = 3.5$ days. Two other spectra were taken within half a day after the first epoch, but these datasets do not carry additional information, thus, we do not include them in our analysis.

The next spectrum was taken at $t_\mathrm{exp} = 13.6$ days, when the photosphere receded to 11\,000 km s$^{-1}$. This agrees with the conclusion of \cite{Zhang22}, where the authors classified SN 2021hpr as a high-velocity gradient \citep[HVG,][]{Benetti05} SN Ia based on the $\sim800$ km s$^{-1}$ day$^{-1}$ decrease of $v_\mathrm{phot}$.

Because of the exponentially decreasing density function toward higher velocities, the dominant light-matter interaction occurs in the few thousand km s$^{-1}$ wide region above the photosphere (except for a few elements like Fe). Thus, the majority of the velocity domain over 10,800 km s$^{-1}$ is poorly sampled. 

Finally, a third epoch at $t_\mathrm{exp} = 19.9$ days is chosen for spectral synthesis ($v_\mathrm{phot}$ = 9,800 km s$^{-1}$). Assuming a linear decrease of $v_\mathrm{phot}$ between the second and third epoch, we characterize $v_\mathrm{phot}$ = 10,000 km s$^{-1}$ at the moment of maximum light.
After the maximum, the assumption of the blackbody emitting photosphere becomes weak in the case of the normal Type Ia SNe, which prevents the computation of realistic spectral fits with {\tt TARDIS}.

\begin{figure}
	\includegraphics[width=\columnwidth]{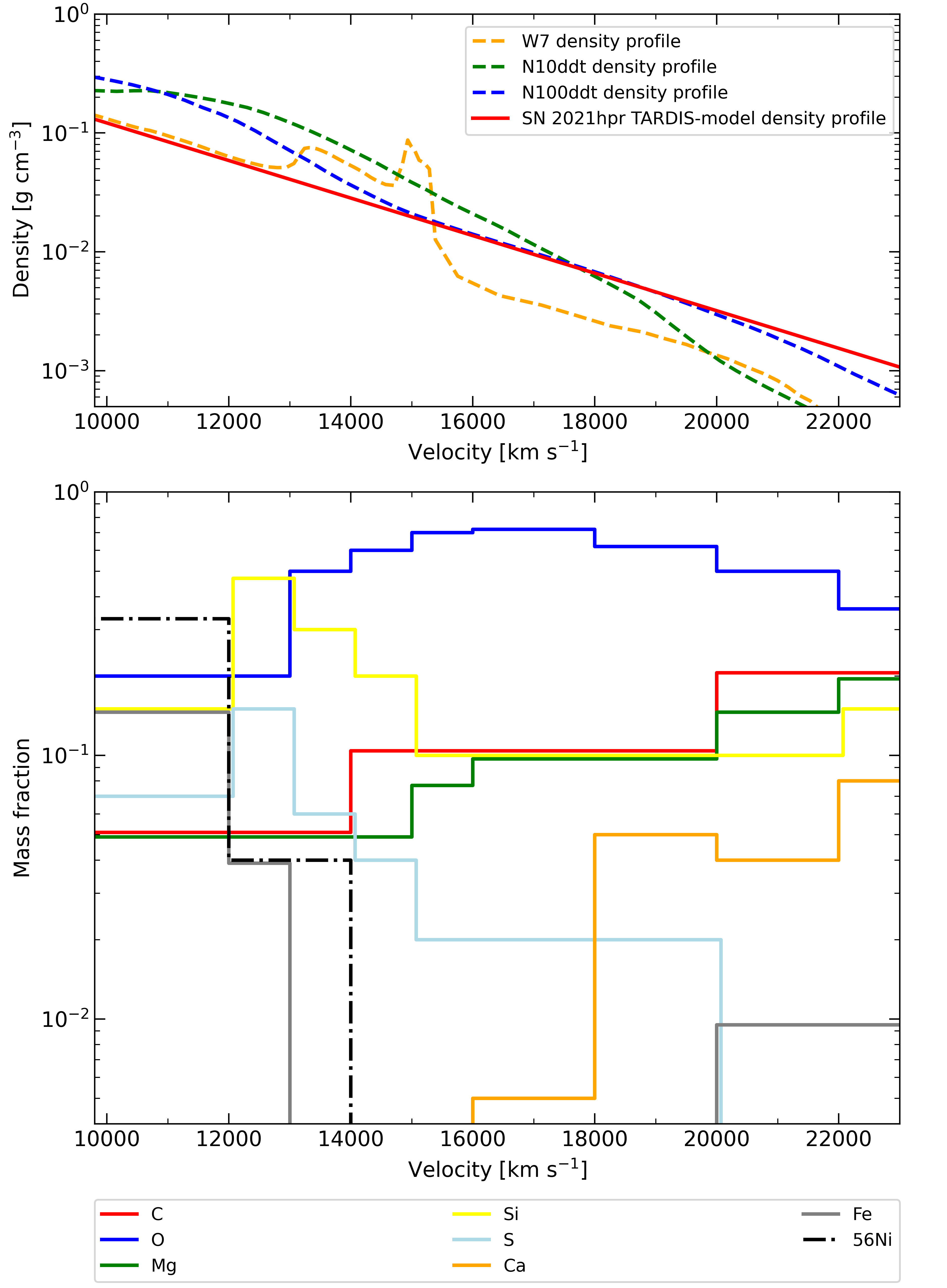}
    \caption{The inferred density (upper panel) and chemical abundances from the abundance tomography carried out via {\tt TARDIS} fits of the three spectral epochs.}
    \label{fig:tardis_densabu_21hpr}
\end{figure}

To reduce the number of the fitting parameters, we set the densities fixed to the exponential fit of the W7 profile (see the upper panel of Fig. \ref{fig:tardis_densabu_21hpr}), adopting $\rho_0 = 4.7$ g\,cm$^{-3}$ as the central density at the reference time $t_0 = 100$ s after the explosion, and $v_\mathrm{0} = 2750$ km\,s$^{-1}$ as the exponential decrease in the function:

\begin{equation}
\rho(t_\mathrm{exp}, v) = \rho_0 \cdot \left(\frac{t_{\mathrm{exp}}}{t_0}\right)^{-3} \cdot \exp\left({-\frac{v}{v_\mathrm{0}}}\right)
\label{eq:fireball2}
\end{equation}

The summary of the input physical parameters producing the well-fitting synthetic spectra of Fig. \ref{fig:tardis_fits_21hpr} can be found in Table \ref{tab:tardis_summary}.

Despite the low time resolution, stratification of the chemical mass fractions can be mapped (Fig. \ref{fig:tardis_densabu_21hpr}), but the steep changes in the abundance functions cannot be tracked due to limited constraints. The first epoch samples the outermost region ($v > 18\,200$ km\,s$^{-1}$), which was designed with the initial assumption of a pure C/O layer. The abundance of C is fit with a monotonously decreasing trend inwards the ejecta to reproduce the C II $\lambda6580$ line. The model of the outermost region also includes increased intermediate-mass element (IME) abundances in order to reproduce the HVFs observed at Ca II H\&K, Si II, Ca II NIR lines frequently reported in the literature. Moreover, Mg is present in the model ejecta with a mass fraction of $X(Mg) > 0.1$, but here the only constraint is the tentative need for the Mg II $\lambda$4481, which, if  real, is severely overlapped with features of ionized iron. The Fe II and Fe III features are very sensitive to the abundance at this early epoch, and we characterize it as $X(Fe) = 0.01$ above 18\,200 km\,s$^{-1}$. All these Fe and IME mass fractions are introduced on the conto of C/O abundances. Note that O is not a well-constrained element is our fitting process, as the O I $\lambda7774$ feature is relatively insensitive to the mass fraction of the element. Thus, we use $X(O)$ as a filler in our chemical composition.

The majority of the model ejecta is designed according to the fit of the second spectral epoch ($v_\mathrm{phot} = 10\,900$ km\,s$^{-1}$), however, some compromise has to be implemented to achieve a better agreement for the third epoch with only a slightly lower photospheric velocity ($v_\mathrm{phot} = 9\,800$ km\,s$^{-1}$). The changes in the abundances of elements can be tracked by the absorption profile of the prominent lines. The red wing of the Ca II H\&K profile caught by the second epoch, constrains the inner abundances of the element with an upper limit of $X(Ca) < 0.0005$ below 16\,000 km\,s$^{-1}$. The Si and S abundances peak around $\sim$13\,000 km\,s$^{-1}$ according to the fit of Si II $\lambda6355$ and S II W feature. The red wings of these absorptions also indicate reduced mass fractions toward the lower velocities.

The IGE elements (except the high-velocity Fe) are limited below $\sim$14\,000 km s$^{-1}$, otherwise, the complex Fe feature around 5000 \r{A} would be excessive at both epochs, especially towards the shorter wavelengths. Due to the limitations of IME abundances (see the paragraph above), Fe and Ni become the dominant elements here.  

Note again that the derived abundance structure should be handled with caution due to the limited fitting constraints provided by the small spectral sample and the high number of fitting parameters. Between 9,800 and 20,000 km\,s$^{-1}$, where we can the following regions can be distinguished in the model ejecta by the general trend of the most prominent elements:
\begin{itemize}
    \item C/O outer region: the chemical profile is dominated by O with an inwards decreasing C contribution, while there is only a moderate IME and almost no IGE abundance.
    \item IME region: the mass fraction of Si higher than 0.5 in a narrow, 1000-2000 km\,s$^{-1}$ wide region; the abundance of C/O drops, while the Ni abundance rises with decreasing velocity.
    \item IGE inner region: the $^{56}$Ni produced in the explosion and its daughter isotopes ($^{56}$Co, $^{56}$Fe) dominate that ejecta
\end{itemize}

This kind of stratification is not specific to only one explosion scenario, instead, most of the deflagration-to-detonation transition and pure detonation models show similar regions with varying locations and relative strengths. The exact velocities, where these regions are separated from each other, just like the chemical abundances, are sensitive to the initial conditions of the hydrodynamic simulations. Thus, a direct quantitative comparison with predictions of any possible explosion scenario is not feasible from the present dataset. 

\begin{table*}
	\centering
	\caption{Summary of input parameters of the {\tt TARDIS} model ejecta for each epochs of SN 2021hpr. The abundances of groups of elements show the average mass fractions within the 2000 km\,s$^{-1}$ range of the photosphere, where the greater part of optical depths emerges. Note that the fitting paramter $L$ indicated here overestimates the bolometric luminosity (see in Sec. \ref{bol_lc}) due to the blackbody assumption of {\tt TARDIS}.}
	\label{tab:tardis_summary}
	\begin{tabular}{|c|c|c|c|c|c|c|} 
    \hline
$t_\mathrm{exp}$ & L & $v_\mathrm{phot}$ & $\rho_\mathrm{phot}$ & X(IGE) & X(IME) & X(C/O) \\
(d) & (10$^{43}$ erg\,s$^{-1}$) & (km\,s$^{-1}$) & (g\,cm$^{-3}$) & \multicolumn{3}{c}{Averages within 2000 km\,s$^{-1}$ above $v_\mathrm{phot}$ } \\
\hline
3.5 & 0.073 & 18.200 & 0.006 & 0.00 & 0.27 & 0.73 \\
13.6 & 1.30 & 10.900 & 0.087 & 0.30 & 0.45 & 0.25 \\
19.9 & 1.05 & 9.800 & 0.130 & 0.48 & 0.27 & 0.25 \\
\hline
\end{tabular}
\end{table*}


\subsubsection{Analysis of the bolometric light curve}
\label{bol_lc}

By using the distance modulus $\mu=33.14$ mag constrained in Sec. \ref{lcfit_21hpr}, physical properties like the initial radioactive nickel mass produced in the explosion (M$_\mathrm{Ni}$) can be constrained. To do so, we construct the pseudo-bolometric light curve of SN 2021hpr using two different methods. In both cases, we apply the available fluxes from Swift UVM2 filter to SDSS z-band, but exclude the Swift UVW1 and UVW2 filters from the process because of their significant red leak \citep{Brown16}. As a first approximation (hereafter referred to as BOL1), we use the SuperBol\footnote{\url{https://zenodo.org/badge/latestdoi/73849147}} code for computing polynomial fits and extrapolations on the individual light curves. Then, before the integration, we extrapolate the observed SEDs with blackbody fits in the bluest and reddest bands for each epoch. By integrating these individual light curves we generate the pseudo-bolometric LC with BB-correction. 

As a second method (BOL2), the flux contribution from the unobserved UV- and IR regimes is estimated in another way. For integrating the UV contribution, the trapezoidal rule is applied with the assumption that the flux reaches zero at 1000 \AA\ \citep{Marion2014, Bora22}. The infrared contribution is taken into account by the exact integration of a Rayleigh-Jeans tail attached to the observed flux at the longest observed wavelength (i.e. $I$- or $i$-band in the present case). Hence, in the following sessions, we refer to this light curve as the pseudo-bolometric LC with Rayleigh-Jeans tail.  

There are multiple ways to estimate the initial $^{56}$Ni mass from the bolometric LC, e.g. the t$_{15}$ method \citep{Sukhbold19} or the tail luminosity method \citep{Afsariardchi21}. Here we fit the estimated pseudo-bolometric LC (supplemented with blackbody corrections)  with a semi-analytic code based on the model of \cite{Arnett82} assuming radiative diffusion in a homologously expanding SN ejecta heated by the radioactive decay of $^{56}$Ni and $^{56}$Co. This model has been developed further by \cite{Valenti08} and \cite{Chatzopoulos12}.

The fitting parameters of the Arnett-model are the mean diffusion timescale, $t_{\mathrm d}$ (also called the light curve timescale, which is practically the geometric mean of the diffusion and the expansion timescales, see \citealt{Arnett82}), the gamma-ray leakage timescale, $T_{\gamma}$. and the initial mass of the radioactive $^{56}$Ni, $M_{\mathrm{Ni}}$. These parameters are directly related to the global physical parameters of the ejecta, namely the total ejecta mass $M_{\mathrm{ej}}$, the characteristic expansion velocity $v_{\mathrm{ exp}}$, the effective optical opacity $\kappa$, and the opacity for gamma-rays $\kappa_{\gamma}$ \citep[see e.g.][]{Konyves-Toth20}.

Since the Arnett-model provides only two timescales ($t_\mathrm{d}$ and $T_\gamma$) for three physical parameters ($M_\mathrm{ej}$, $v_\mathrm{exp}$ and $\kappa$), the physical parameters cannot be constrained independently from only photometry. To overcome this difficulty, \citet{Konyves-Toth20} used an approximate, iterative method by estimating a lower and an upper limit for the optical opacity first, then using the average of them to constrain the ejecta mass and the expansion velocity.


Recently the diffusion model was challenged by \cite{Khatami19}, who suggested an alternative formalism to estimate the initial nickel masses for various types of SN explosions. However, \cite{Bora22} demonstrated that for Type Ia SNe in particular, the nickel masses inferred from the two methods are consistent within their uncertainties. 

\begin{figure*}
	\centering
	\includegraphics[width=15cm]{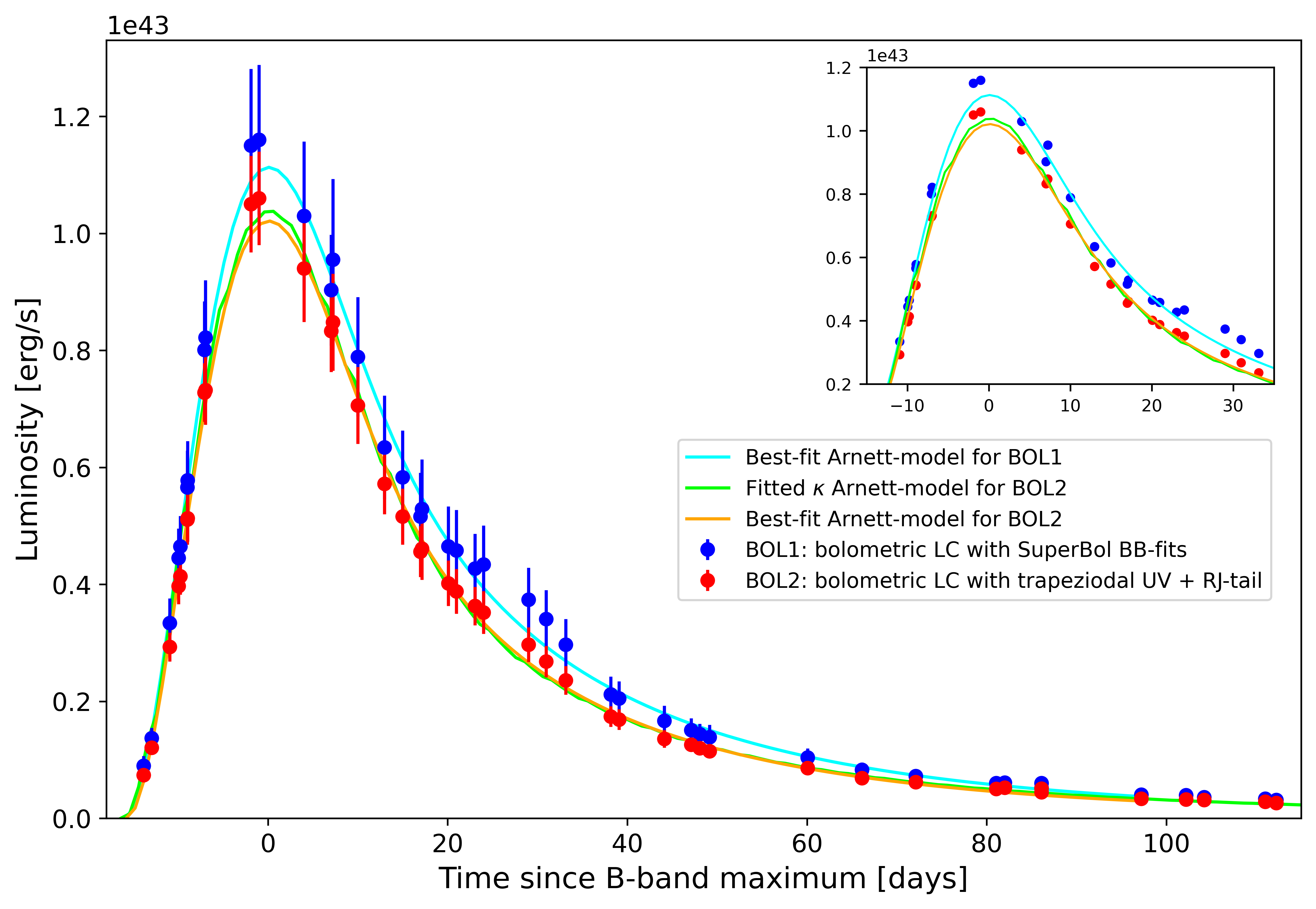}
    \caption{ Pseudo-bolometric light curves estimated from the integration of observed magnitudes with trapezoidal approximation for the UV and Rayleigh-Jeans tail for the NIR contribution (red), and the LC estimated by blackbody fits to the UV/NIR regions (blue), and their best-fit Arnett-models (see Sec. \ref{bol_lc}.)}
    \label{fig:arnett}
\end{figure*}

As a first attempt, we fit the pseudo-bolometric light curve (see in Fig. \ref{fig:arnett}) provided by SuperBol (BOL1). The best-fit model parameters estimated by the {\tt Minim} code \citep{Chatzopoulos13} are shown in Table~\ref{tab:fitting_data} together with the inferred physical parameters. These data 
suggest a nickel mass of $M_\mathrm{Ni} \sim 0.48$ M$_\odot$, which is consistent with the absolute peak magnitude of $M_\mathrm{max}(B) \sim 19$ mag. 
However, the BOL1 LC suffers from higher uncertainties mainly due to the imperfect BB-extensions to the near-UV and near-IR regime fit by the SuperBol code. The strong bump between +20 and +30 days is partially non-physical as the BB-fits of the code overestimate the flux contribution from the not-observed spectral regimes. Since these uncertainties lead to an inferior fitting, we disfavour the results from the BOL1 LC.

The BOL2 light curve is also fit (Fig. \ref{fig:arnett}) following the same methodology, and the inferred physical quantities from the best-fit parameters (see Table~\ref{tab:fitting_data} indicate a realistic M$_\mathrm{Ni} \sim 0.44$ M$_\odot$, and $v_\mathrm{exp} \sim 11200$ km s$^{-1}$. 
The mean optical opacity is constrained as $\kappa \sim 0.144$ according to the iterative method by \cite{Konyves-Toth20}, which corresponds to an ejecta mass of $\sim 1.12$ M$_\odot$. 

As an alternative solution, the same set of best-fit parameters can be used with a fixed expansion velocity adopted from the spectral analysis (Section~\ref{tardis}). The results are shown in the 4th column of Table~\ref{tab:fitting_data}. By assuming $v_\mathrm{phot}$ = 10\,000 km s$^{-1}$ as v$_\mathrm{exp}$, we get a lower ejecta mass of $M_\mathrm{ej} \sim 0.89$ M$_\odot$. Note, however, that even though $v_\mathrm{phot}$ is generally used as an approximation for $v_\mathrm{exp}$, these two velocities are not the same quantity by definition \citep[see][]{Arnett82}. Thus, this estimate can be considered only as a lower limit for the ejecta mass.

For further validation, we can also constrain the value of the effective optical opacity ($\kappa$) directly from fitting the light curve with the Arnett-model-based LC2 code \citep{Nagy16} coupled with {\tt Minim}. In LC2 the ejecta mass, the radius of the progenitor, the nickel mass, the opacity and the kinetic energy of the ejecta can be chosen as fitting parameters. Since our data span only the first few months after the explosion, we ignore the leaking of positrons from the ejecta, and assume only the usual gamma-ray leaking with an effective gamma-ray opacity of $\kappa_\gamma \sim 0.03$ cm$^2$g$^{-1}$. The advantage of this approach is that the highly uncertain expansion velocity is not a fitting (and neither an input) parameter, however, the results are sensitive to the value of $\kappa$ included in the fitting. Note, that due to significant parameter correlations in the Arnett-model \citep[e.g.][]{Nagy14, Nagy16} all the previously described methods are tainted with significant systematic uncertainties. We take into account these correlations while estimating the average errors of the inferred physical properties of SN 2021hpr (see Table \ref{tab:fitting_data}). 

The main parameters inferred from the fitted $\kappa$ method ($M_\mathrm{ej} \sim 1.28$ M$_\odot$) are consistent with those estimated by the previous approaches and are rather close to the results of the mean $\kappa$ method. Thus, we accept $M_\mathrm{Ni} = 0.44 \pm 0.14$ M$_\odot$ and the other corresponding parameters (see the 3rd column with boldface fonts in Table \ref{tab:fitting_data}) as the final result of the bolometric LC analysis.


  \begin{table*}[!h]
	\centering
	\caption{Comparison of the physical parameters of SN 2021hpr from fitting the pseudo-bolometric light curves, taking into account gamma-ray leakage. The three methods are a) assuming the average of the upper and lower limit of $\kappa$ as effective opacity \citep[mean $\kappa$,][]{Konyves-Toth20}; b) assuming a fixed expansion velocity from the spectroscopic analysis \citep[adopted $v_\mathrm{exp}$,][]{Konyves-Toth20}; c) direct fit of the effective opacity \citep[fitted $\kappa$,][]{Nagy14}. The approximate uncertainties are computed from error propagation in the corresponding method. The only exception is $M_\mathrm{Ni}$, where the {\tt Minim} algorithm provided too low uncertainties from the fits; therefore, the uncertainty of the nickel masses is estimated as in \cite{Nagy16}. }
	\label{tab:fitting_data}
	\begin{tabular}{l|c|cc|c} 
		\hline
		  Bolometric light curve & BOL1 & \multicolumn{3}{c}{BOL2} \\
            \hline
            \hline
            Parameter  & mean $\kappa$ & \textbf{mean $\kappa$} & adopted $v_\mathrm{exp}$ & fitted $\kappa$\\ 
		\hline
		  $M_\mathrm{Ni}$ [M$_\odot$] & 0.48 (0.16) & \textbf{0.44 (0.14)} & 0.44 (0.14) & 0.46 (0.15) \\
            $t_\mathrm{d}$ [day] & 13.72 (0.30) & \textbf{13.68 (0.227)} & 13.68 (0.227) & 16.3 (1.22) \\
            $T_\mathrm{\gamma}$ [day] & 44.83 (0.84) & \textbf{41.47 (0.58)} & 41.47 (0.58) & 50.1 (4.6) \\
            \hline
            $\kappa$ [cm$^2$ g$^{-1}$] & 0.139 (0.010) & \textbf{0.144 (0.017)} & 0.161 (0.018) & 0.16 (0.01) \\
            $v_\mathrm{exp}$ [$10^{3}$ km\,s$^{-1}$]  & 10.8 (1.0) & \textbf{11.2 (1.2)} & 10.0 (1.0) & 9.9 (0.7) \\
            M$_\mathrm{ej}$ [M$_\odot$] & 1.22 (0.24) & \textbf{1.12 (0.28)} & 0.89 (0.21) & 1.28 (0.15) \\
        E$_{kin}$ [$10^{51}$ erg] & 0.85 (0.38) & \textbf{0.84 (0.29)} & 0.54 (0.28) & 0.75 (0.10) \\
		\hline
	\end{tabular}
\end{table*}



\section{Conclusion}
\label{conclusion}

We analyzed the optical/UV light curve of the normal Type Ia SN 2021hpr together with its two well-observed siblings, SNe 1997bq and 2008fv, that appeared in the same host galaxy, NGC~3147. The three SNe provide a unique opportunity to revise the distance of their host, and allow the testing of systematic effects influencing the various distance estimation methods.

We took new photometric data on SN~2021hpr with the twin robotic telescopes, RC80 and BRC80, from two Hungarian observatories (Konkoly and Baja). 
The light curves, including $BVgriz$ filters, start from $t=\, -14.9$ days, adopting 59323.0 MJD as the moment of maximum light in the $B$-band. Based on fitting the early data points, we constrained $T_{first} \sim 59304.4$ MJD as the date of first light, which gives $t_\mathrm{rise} = 18.4$ days for the rise time. Besides the optical follow-up, we also downloaded and analyzed the available Swift UVOT photometry taken from $-16.1$ to $+12.3$ days relative to the $B$-band maximum.

We estimated the distance to SN~2021hpr by fitting our new $BVgri$ data with the LC-fitter codes MLCS2k2, SALT2 and SALT3. The same three codes were applied for the two sibling SNe, and the inferred individual distances were compared to each other. We found that the results scatter between $\mu = 32.97$ and $33.35$ mag with a mean value of $\mu = 33.12 \pm 0.16$ mag. This estimation for the distance of NGC 3147 is in good agreement with previous SN-based studies (see Table~\ref{tab:distance_summary} and Figure \ref{fig:dist_moduli}), and also marginally consistent with the most recent Cepheid-based distance \citep[$\mu_\mathrm{Ceph}=33.01 \pm 0.165$ mag,][]{Riess22}.

We adopted the distance inferred from the SALT3 fitting to SN~2021hpr, where the most complete self-consistent optical dataset was applied and the models provided good fits to all bands. The SALT3-based distance modulus $\mu = 33.14 \pm 0.05$ mag is also consistent with that of \cite{Ward22} estimated  by the BayeSN code from fitting to an independent LC of SN 2021hpr. 

In order to study the physical parameters of SN~2021hpr based on our new, improved distance, abundance tomography was performed by fitting three pre-maximum spectra with the radiative transfer code {\tt TARDIS}. The constrained chemical structure is consistent with multiple explosion models, including deflagration-to-detonation transition and pure detonation scenarios, with an additional overabundance of Mg, Si and Ca to reproduce the observed blue-shifts and high-velocity features of the corresponding absorption lines. 

We calculated the bolometric LC of SN~2021hpr from its optical data supplemented by near-UV data from Swift, and fit it with the radiative diffusion Arnett-model. The peak luminosity turned out to be $1.02 \times 10^{43}$ erg/s, while the total mass of $^{56}$Ni produced in the explosion is estimated as $0.44 \pm 0.14 M_\odot$, while the ejecta mass and expansion velocity were calculated as $1.12 \pm 0.28$ M$_\odot$ and $11\,200 \pm 1200$ km~s$^{-1}$, respectively. Considering all the estimated physical properties and spectroscopic characteristics of SN 2021hpr, it belongs to the Branch-normal and high velocity gradient classes of Type Ia SNe. 

\begin{acknowledgements}
This research made use of \textsc{tardis}, a community-developed software package for spectral synthesis in supernovae \citep{Kerzendorf14, kerzendorf_wolfgang_2023_7855824}. The
development of \textsc{tardis} received support from GitHub, the Google Summer of Code
initiative, and from ESA's Summer of Code in Space program. \textsc{tardis} is a fiscally
sponsored project of NumFOCUS. \textsc{tardis} makes extensive use of Astropy and Pyne.
The authors acknowledge the Hungarian National Research, Development and Innovation Office
grants OTKA K-131508, K-138962, K-142534, FK-134432, KKP-143986 (\'Elvonal), and 2019-2.1.11-T\'eT-2019-00056. LK acknowledges the Hungarian National Research, Development and Innovation Office grant OTKA PD-134784. 
BB, RKT and ZsB is supported by the ÚNKP-22-2 New National Excellence Program of the Ministry for Culture and Innovation from the source of the National Research, Development and Innovation Fund.
APN is supported by NKFIH/OTKA PD-134434 grant, which is founded by the Hungarian National Development and Innovation Fund.  LK, KV, and TS are Bolyai J\'anos Research Fellows of the Hungarian Academy of Sciences. KV and TS are supported by the Bolyai+ grants \'UNKP-22-5-ELTE-1093 and \'UNKP-22-5-SZTE-591, respectively. SZ is supported by the National Talent Programme under NTP-NFTÖ-22-B-0166 Grant. ZMS acknowledges funding from a St Leonards scholarship from the University of St Andrews. ZMS is a member of the International Max Planck Research School (IMPRS) for Astronomy and Astrophysics at the Universities of Bonn and Cologne.
\end{acknowledgements}

%
%
\bibliographystyle{aa}
\bibliography{example} 

\clearpage

\begin{appendix} 

\onecolumn
\section{Observational data}

\begin{center}
\begin{longtable}{ccccccc}
\caption{Log of SN 2021hpr photometry from Observatory of Baja. } \label{tab:photometric_data_BRC80} \\
\hline
\multicolumn{1}{c}{\textbf{MJD}} & \multicolumn{1}{c}{\textbf{B}} & \multicolumn{1}{c}{\textbf{V}} & \multicolumn{1}{c}{\textbf{g}} & \multicolumn{1}{c}{\textbf{r}} & \multicolumn{1}{c}{\textbf{i}} & \multicolumn{1}{c}{\textbf{z}} \\ \hline

59312.1	 & 	15.312 (0.08) 	 & 	15.237 (0.06) 	 & 	15.201 (0.01) 	 & 	15.201 (0.01) 	 & 	15.509 (0.01) 	 & 	15.528 (0.02) \\
59312.9	 & 	15.050 (0.16) 	 & 	15.022 (0.08) 	 & 	15.004 (0.01) 	 & 	15.012 (0.01) 	 & 	15.324 (0.01) 	 & 	15.398 (0.02) \\
59314.8 	 & 	14.684 (0.06) 	 & 	14.684 (0.04) 	 & 	14.671 (0.01) 	 & 	14.733 (0.02) 	 & 	14.998 (0.01) 	 & 	15.070 (0.02) \\
59329.1 	 & 	14.418 (0.10) 	 & 	14.266 (0.05) 	 & 	14.327 (0.01) 	 & 	14.321 (0.01) 	 & 	15.045 (0.01) 	 & 	15.085 (0.04) \\
59339.0 	 & 	15.481 (0.22) 	 & 	14.762 (0.05) 	 & 	14.948 (0.03) 	 & 	14.893 (0.02) 	 & 	15.484 (0.03) 	 & 	15.192 (0.10) \\
59403.8 	 & 	17.678 (0.34) 	 & 	17.114 (0.08) 	 & 	17.331 (0.05) 	 & 	17.119 (0.03) 	 & 	17.610 (0.04) 	 & 	18.012 (0.14) \\
59407.9 	 & 	17.700 (0.07) 	 & 	17.174 (0.03) 	 & 	17.350 (0.02) 	 & 	17.242 (0.02) 	 & 	17.600 (0.03) 	 & 	17.871 (0.11) \\
59426.0 	 & 	18.023 (0.37) 	 & 	17.675 (0.09) 	 & 	17.708 (0.07) 	 & 	17.916 (0.05) 	 & 	18.222 (0.06) 	 & 	18.631 (0.22) \\
59432.8 	 & 	18.269 (0.95) 	 & 	17.672 (0.10) 	 & 	17.789 (0.04) 	 & 	17.978 (0.98) 	 & 	18.403 (0.06) 	 & 	18.720 (0.16) \\
59446.8 	 & 	18.362 (0.29) 	 & 	18.062 (0.08) 	 & 	18.109 (0.07) 	 & 	18.398 (0.06) 	 & 	18.748 (0.16) 	 & 	18.209 (0.29) \\
59459.7 	 & 	18.356 (0.78) 	 & 	18.313 (0.13) 	 & 	18.112 (0.05) 	 & 	18.793 (0.06) 	 & 	18.989 (0.10) 	 & 	19.095 (0.30) \\
59467.8 	 & 	18.854 (0.15) 	 & 	18.433 (0.11) 	 & 	18.349 (0.05) 	 & 	19.224 (0.08) 	 & 	19.347 (0.12) 	 & 	18.937 (0.27) \\

\hline
\hline
\end{longtable}
\end{center}

\begin{center}
\begin{longtable}{ccccccc}
\caption{Log of SN 2021hpr photometry from Observatory of Piszkesteto. } \label{tab:photometric_data_PSCH80} \\
\hline
\multicolumn{1}{c}{\textbf{MJD}} & \multicolumn{1}{c}{\textbf{B}} & \multicolumn{1}{c}{\textbf{V}} & \multicolumn{1}{c}{\textbf{g}} & \multicolumn{1}{c}{\textbf{r}} & \multicolumn{1}{c}{\textbf{i}} & \multicolumn{1}{c}{\textbf{z}} \\ \hline

59308.01	&	17.72 (0.10)	&	16.78 (0.06)	&	17.06 (0.05)	&	16.81 (0.04)	&	17.27 (0.07)	&	16.90 (0.13)	\\
59308.91	&	17.00 (0.10)	&	16.31 (0.04)	&	16.49 (0.03)	&	16.36 (0.03)	&	16.74 (0.03)	&	16.50 (0.06)	\\
59310.93	&	15.83 (0.08)	&	15.45 (0.05)	&	15.52 (0.02)	&	15.53 (0.02)	&	15.87 (0.02)	&	15.81 (0.04)	\\
59311.90	&	15.45 (0.09)	&	15.15 (0.05)	&	15.27 (0.04)	&	15.23 (0.03)	&	15.57 (0.02)	&	15.61 (0.04)	\\
59312.92	&	15.19 (0.07)	&	14.93 (0.04)	&	14.98 (0.03)	&	15.00 (0.02)	&	15.31 (0.02)	&	15.35 (0.04)	\\
59314.90	&	14.84 (0.08)	&	14.62 (0.04)	&	14.63 (0.02)	&	14.66 (0.02)	&	15.00 (0.02)	&	15.12 (0.04)	\\
59319.97	&	14.45 (0.09)	&	14.20 (0.04)	&	14.24 (0.03)	&	14.27 (0.03)	&	14.83 (0.02)	&	14.89 (0.04)	\\
59320.87	&	14.42 (0.08)	&	14.18 (0.04)	&	14.22 (0.03)	&	14.24 (0.03)	&	14.83 (0.02)	&	14.89 (0.04)	\\
59324.79	&	14.46 (0.08)	&	14.13 (0.04)	&	13.95 (0.09)	&	14.17 (0.04)	&	14.90 (0.03)	&	14.91 (0.05)	\\
59325.88	&	14.49 (0.20)	&	14.21 (0.05)	&	14.29 (0.07)	&	14.27 (0.02)	&	14.92 (0.03)	&	14.97 (0.05)	\\
59328.89	&	14.61 (0.16)	&	14.30 (0.05)	&	14.34 (0.09)	&	14.30 (0.05)	&	15.06 (0.04)	&	15.10 (0.05)	\\
59331.88	&	14.83 (0.13)	&	14.35 (0.05)	&	14.48 (0.05)	&	14.49 (0.03)	&	15.22 (0.04)	&	15.17 (0.05)	\\
59334.84	&	15.11 (0.08)	&	14.57 (0.04)	&	14.68 (0.04)	&	14.71 (0.03)	&	15.45 (0.03)	&	15.16 (0.09)	\\
59335.99	&	15.28 (0.15)	&	14.65 (0.06)	&	14.77 (0.04)	&	14.80 (0.04)	&	15.47 (0.03)	&	--	\\
59336.84	&	15.25 (0.10)	&	14.69 (0.04)	&	14.83 (0.04)	&	14.81 (0.03)	&	15.47 (0.03)	&	15.14 (0.08)	\\
59338.85	&	15.56 (0.09)	&	14.79 (0.04)	&	14.98 (0.03)	&	14.87 (0.03)	&	15.49 (0.02)	&	15.23 (0.21)	\\
59341.94	&	15.90 (0.12)	&	14.97 (0.04)	&	15.22 (0.03)	&	14.93 (0.03)	&	15.43 (0.02)	&	15.08 (0.04)	\\
59342.85	&	16.00 (0.09)	&	15.01 (0.05)	&	15.31 (0.03)	&	14.94 (0.03)	&	15.41 (0.02)	&	15.08 (0.05)	\\
59344.92	&	16.21 (0.08)	&	15.11 (0.03)	&	15.48 (0.03)	&	14.96 (0.03)	&	15.36 (0.02)	&	15.05 (0.05)	\\
59345.86	&	16.28 (0.09)	&	15.17 (0.03)	&	15.55 (0.02)	&	14.98 (0.02)	&	15.34 (0.02)	&	15.07 (0.05)	\\
59350.87	&	16.69 (0.08)	&	15.44 (0.04)	&	15.95 (0.03)	&	15.13 (0.03)	&	15.32 (0.02)	&	15.07 (0.05)	\\
59352.83	&	16.86 (0.13)	&	15.57 (0.05)	&	16.09 (0.03)	&	15.26 (0.02)	&	15.42 (0.02)	&	15.11 (0.04)	\\
59354.92	&	16.98 (0.08)	&	15.73 (0.04)	&	16.24 (0.04)	&	15.41 (0.03)	&	15.54 (0.03)	&	15.25 (0.04)	\\
59359.94	&	17.34 (0.15)	&	16.04 (0.06)	&	16.50 (0.07)	&	15.72 (0.03)	&	15.90 (0.02)	&	15.62 (0.05)	\\
59360.88	&	17.42 (0.16)	&	16.04 (0.06)	&	16.49 (0.09)	&	15.73 (0.04)	&	15.94 (0.02)	&	15.67 (0.06)	\\
59365.91	&	17.46 (0.08)	&	16.19 (0.06)	&	16.66 (0.04)	&	15.99 (0.04)	&	16.27 (0.04)	&	15.99 (0.08)	\\
59368.91	&	17.34 (0.08)	&	16.28 (0.04)	&	16.71 (0.04)	&	16.09 (0.03)	&	16.36 (0.02)	&	16.17 (0.06)	\\
59369.84	&	17.40 (0.16)	&	16.37 (0.06)	&	16.78 (0.05)	&	16.14 (0.05)	&	16.39 (0.03)	&	16.17 (0.08)	\\
59370.95	&	17.51 (0.10)	&	16.38 (0.05)	&	16.78 (0.04)	&	16.18 (0.03)	&	16.47 (0.03)	&	16.28 (0.08)	\\
59381.86	&	17.55 (0.13)	&	16.67 (0.05)	&	16.98 (0.05)	&	16.54 (0.03)	&	16.92 (0.05)	&	16.64 (0.07)	\\
59387.90	&	17.82 (0.14)	&	16.82 (0.05)	&	17.23 (0.07)	&	16.80 (0.04)	&	17.13 (0.05)	&	17.14 (0.11)	\\
59387.90	&	17.82 (0.14)	&	16.82 (0.05)	&	17.23 (0.07)	&	16.80 (0.04)	&	17.13 (0.05)	&	17.14 (0.11)	\\
59393.88	&	17.74 (0.10)	&	16.97 (0.04)	&	17.20 (0.03)	&	16.93 (0.03)	&	17.37 (0.03)	&	17.29 (0.11)	\\
59402.86	&	17.82 (0.11)	&	17.17 (0.04)	&	17.38 (0.04)	&	17.16 (0.02)	&	17.61 (0.04)	&	17.85 (0.17)	\\
59407.88	&	17.95 (0.09)	&	17.27 (0.04)	&	17.43 (0.03)	&	17.36 (0.03)	&	17.80 (0.03)	&	17.91 (0.09)	\\
59419.01	&	18.25 (0.23)	&	17.58 (0.11)	&	17.66 (0.12)	&	17.75 (0.07)	&	18.06 (0.11)	&	18.45 (0.35)	\\
59423.97	&	18.13 (0.09)	&	17.59 (0.04)	&	17.75 (0.04)	&	17.82 (0.04)	&	18.21 (0.08)	&	18.58 (0.24)	\\
59429.04	&	17.95 (0.25)	&	17.76 (0.10)	&	18.32 (0.32)	&	18.08 (0.09)	&	18.20 (0.32)	&	--	\\
59434.02	&	18.36 (0.08)	&	17.84 (0.05)	&	17.81 (0.04)	&	18.11 (0.03)	&	18.60 (0.11)	&	19.28 (0.62)	\\

\hline
\hline
\end{longtable}
\end{center}

\begin{center}
\begin{longtable}{ccccccc}
\caption{Log of Swift UVOT photometry of SN 2021hpr. } \label{tab:photometric_data_Swift} \\
\hline
\multicolumn{1}{c}{\textbf{MJD}} & \multicolumn{1}{c}{\textbf{UW2}} & \multicolumn{1}{c}{\textbf{UM2}} & \multicolumn{1}{c}{\textbf{UW1}} & \multicolumn{1}{c}{\textbf{U}} & \multicolumn{1}{c}{\textbf{B}} & \multicolumn{1}{c}{\textbf{V}} \\ \hline  
59335.240 & 18.005 (0.102) & 18.741 (0.159) & 16.788 (0.059) & 15.379 (0.033) & 15.062 (0.020) & 14.553 (0.026) \\ 
59332.190 & 17.723 (0.084) & 18.390 (0.125) & 16.354 (0.045) & 15.030 (0.027) & 14.770 (0.017) & 14.401 (0.024) \\ 
59331.060 & 17.875 (0.191) & 18.785 (0.358) & 16.249 (0.087) & 14.944 (0.053) & 14.544 (0.033) & 14.292 (0.047) \\ 
59326.290 & 17.178 (0.079) & 18.192 (0.150) & 15.848 (0.047) & 14.284 (0.025) & 14.293 (0.020) & 14.085 (0.028) \\ 
59321.590 & -- & -- & 15.552 (0.026) & -- & -- & -- \\ 
59321.570 & 17.084 (0.062) & -- & -- & -- & -- & -- \\ 
59318.730 & 17.014 (0.059) & 18.414 (0.154) & 15.445 (0.032) & 13.882 (0.018) & 14.220 (0.016) & 14.255 (0.026) \\ 
59316.660 & 17.206 (0.060) & 18.509 (0.146) & 15.633 (0.030) & 14.031 (0.017) & 14.326 (0.015) & 14.415 (0.025) \\ 
59311.890 & 18.208 (0.260) & 19.094 (0.564) & 16.925 (0.137) & 15.358 (0.066) & 15.251 (0.045) & 15.184 (0.079) \\ 
59309.830 & -- & -- & 18.046 (0.161) & 16.727 (0.063) & 16.115 (0.031) & 15.754 (0.067) \\ 
59308.460 & -- & -- & 18.651 (0.266) & 17.704 (0.128) & 17.010 (0.054) & 16.438 (0.091) \\ 
59307.240 & -- & -- & 18.983 (0.353) & 18.217 (0.195) & 17.753 (0.096) & 17.503 (0.244) \\ 
59306.800 & -- & 18.759 (0.346) & 18.180 (0.240) & 17.320 (0.161) & 18.078 (0.277) & -- \\
\hline

\end{longtable}
\end{center}

\begin{center}
\begin{longtable}{ccccc} 
\caption{Log of the spectra; $t_\mathrm{exp}$ shows the time since the date of explosion constrained in the abundance tomography (MJD $59\,304.0$); while the phases are given relative to the maximum in $B$-band (MJD $59\,321.9$). } \label{tab:spectroscopic_data} \\
\hline
\multicolumn{1}{c}{\textbf{MJD}} & \multicolumn{1}{c}{\textbf{t$_\mathrm{exp}$ [d]}} & \multicolumn{1}{c}{\textbf{Phase [d]}} & \multicolumn{1}{c}{\textbf{Telescope/Instrument}} & \multicolumn{1}{c}{\textbf{Wavelength range [\r{A}]}} \\ \hline 
		59307.5 & 3.5 & -14.4 & XLT/BFOSC & 3700 - 8800\\
		59317.6 & 13.6 & -4.3 & XLT/BFOSC & 3700 - 8800\\
		59323.9 & 19.9 & +2.0 & Smolecin Observatory & 3900 - 7100\\
		\hline

\end{longtable}
\end{center}

\section{Light curve fits}

In the followings, the plots of MLCS2k2, SALT2.4 and SALT3 fits for SNe 1997bq, 2008fv and 2021hpr are listed. The distance moduli estimated from these fits are discussed in Sec. \ref{distance}, while the corresponding fitting parameters are listed in Tab. \ref{tab:fitting_params}.

\begin{center}
\label{tab:fitting_params}
\begin{longtable}{|l|ccc|ccc|cc|} 
\caption{Log of the light curve fits. }
\label{tab:lc_fits} \\
 \hline
 & \multicolumn{3}{c}{\textbf{SALT3}}  & \multicolumn{3}{c}{\textbf{SALT2.4}}  & \multicolumn{2}{c}{\textbf{MLCS2k2}} \\
 \hline
 & x$_0$ & x$_1$ & c  & x$_0$ & x$_1$ & c & A$_\mathrm{host}$ & $\Delta$\\ 
 \hline
SN 1997bq & 0.03142 & -1.025 & 0.0696 & 0.0305 & -0.6073 & 0.1094 & 0.60 & 0.00 \\
SN 2008fv & 0.0260 & 0.9040 & 0.1452 & 0.02688 & 0.7690 & 0.1433 & 0.90 & -0.29  \\
SN 2021hpr & 0.04091 & -0.044 & 0.0051 & 0.03739 & 0.4483 & 0.0423 & 0.45 & 0.03  \\
\hline
\end{longtable}
\end{center}

\pagebreak

\twocolumn

\begin{figure}
	\includegraphics[width=100mm]{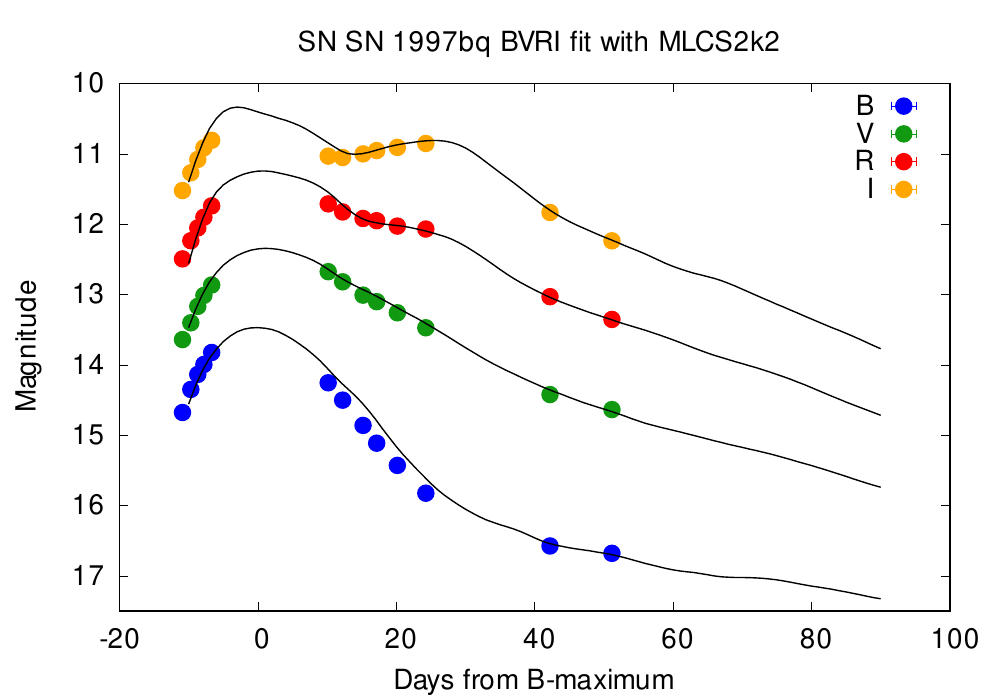}
    \caption{MLCS2k2 LC model fitting of the photometry of SN 1997bq}
    \label{fig:mlcs_97bq}
\end{figure}

\begin{figure}
	\includegraphics[width=100mm]{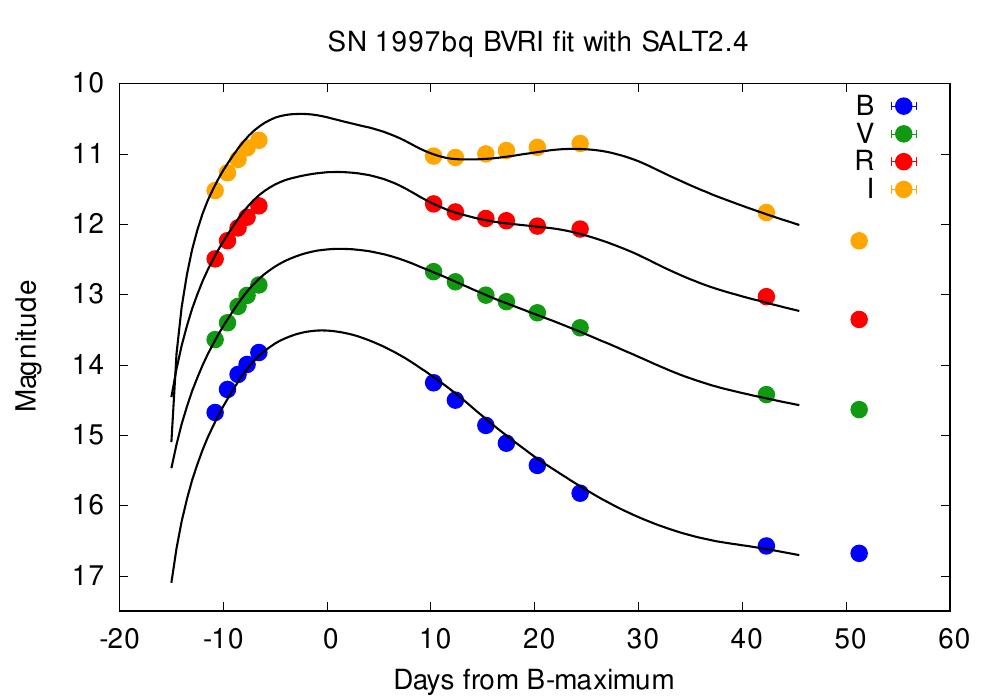}
    \caption{SALT2.4 LC model fitting of the photometry of SN 1997bq}
    \label{fig:salt2_97bq}
\end{figure}

\begin{figure}
	\includegraphics[width=100mm]{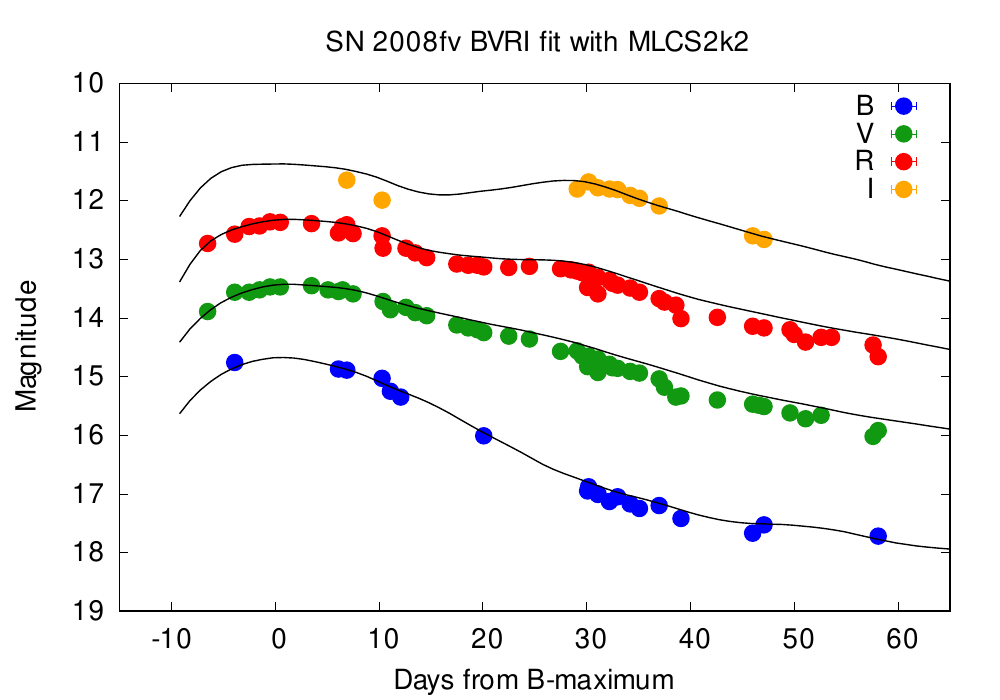}
    \caption{MLCS2k2 LC model fitting of the photometry of SN 2008fv.}
    \label{fig:mlcs_08fv}
\end{figure}

\begin{figure}
	\includegraphics[width=100mm]{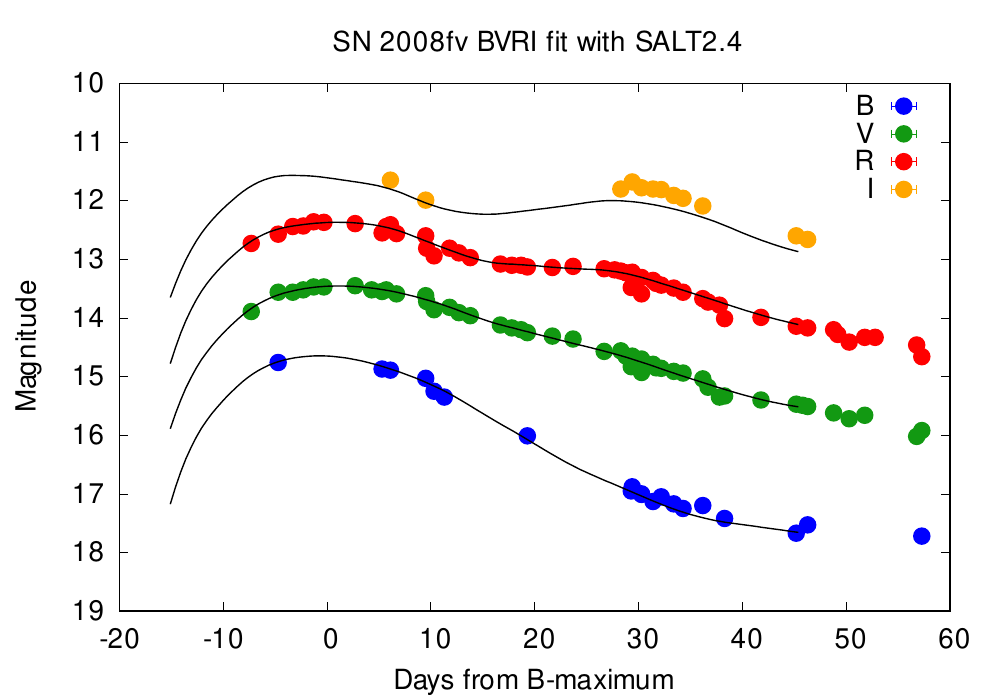}
    \caption{SALT2.4 LC model fitting of the photometry of SN 2008fv.}
    \label{fig:salt2_08fv}
\end{figure}

\begin{figure}
	\includegraphics[width=100mm]{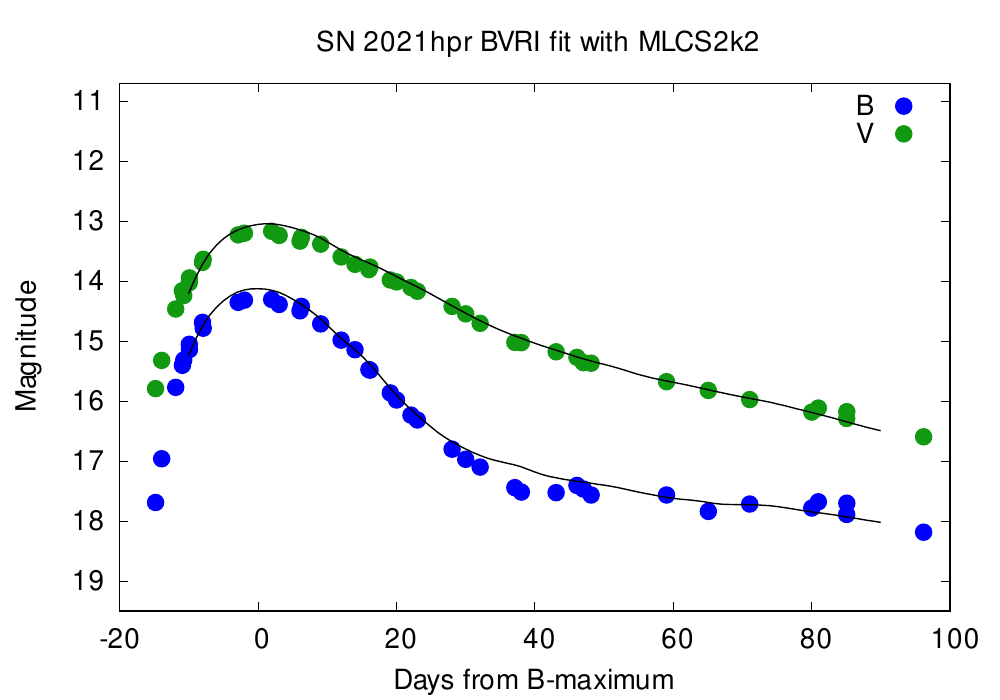}
    \caption{MLCS2k2 LC model fitting of the photometry of SN 2021hpr.}
    \label{fig:mlcs_21hpr4}
\end{figure}

\begin{figure}
	\includegraphics[width=100mm]{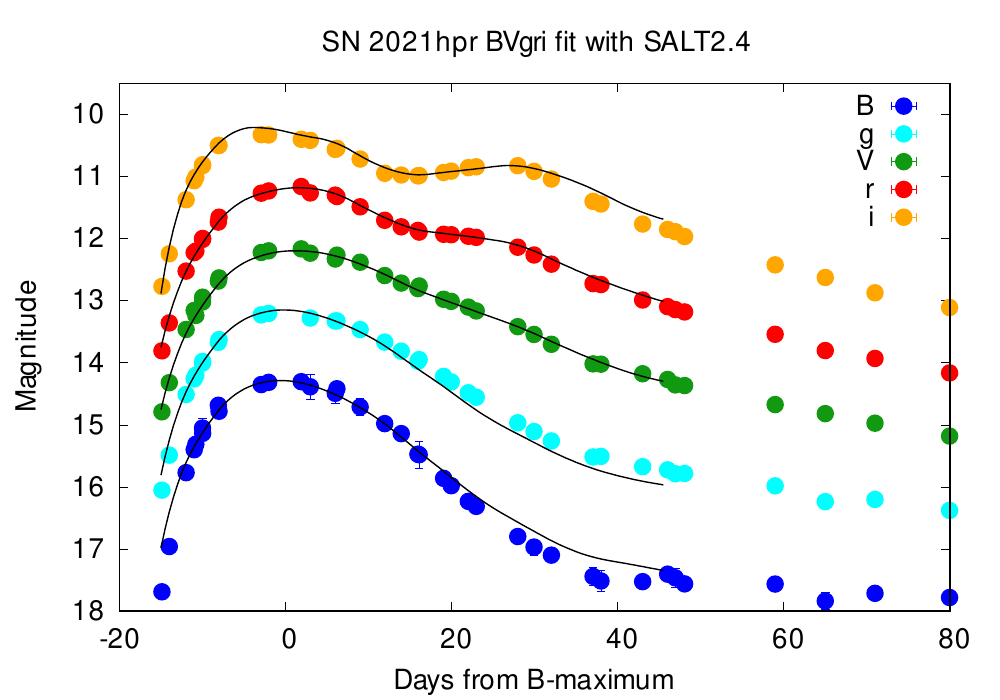}
    \caption{SALT2.4 LC model fitting of the photometry of SN 2021hpr.}
    \label{fig:salt2_21hpr5}
\end{figure}

\begin{figure}
	\includegraphics[width=100mm]{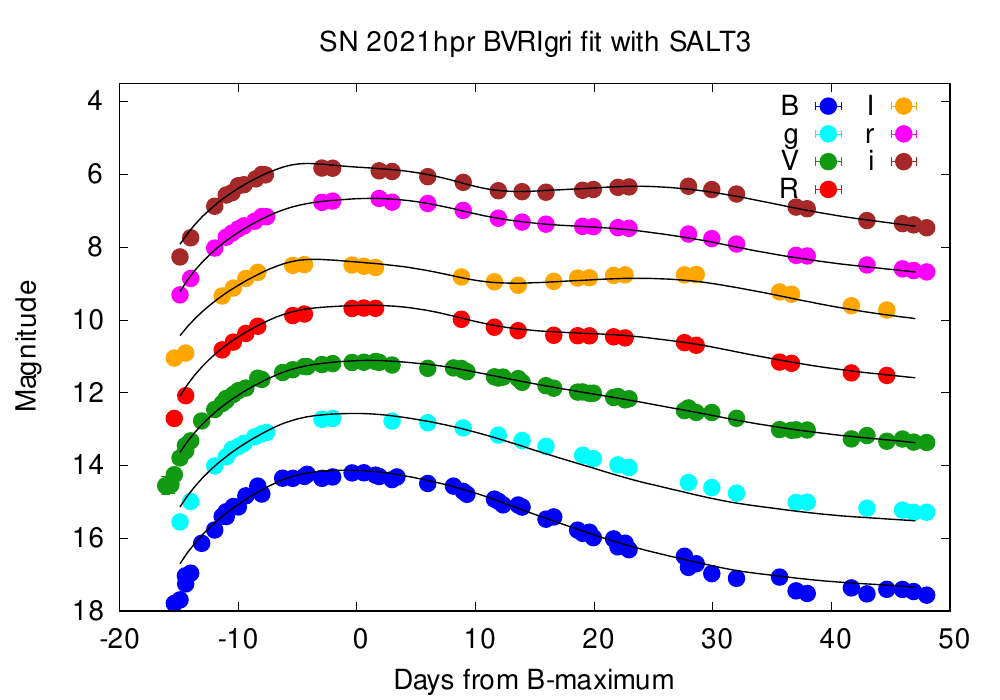}
    \caption{SALT3 LC model fitting of the extended photometric dataset of SN 2021hpr.}
    \label{fig:salt3_21hpr6}
\end{figure}

\end{appendix}

\end{document}